\documentclass[twocolumn,superscriptaddress,showpacs,preprintnumbers,amsmath,amssymb]{revtex4}
\usepackage{amsmath}
\usepackage{amssymb}
\usepackage{dcolumn}
\usepackage{graphicx}
\usepackage{bm}
\usepackage{epstopdf}
\usepackage{threeparttable}
\usepackage{float}
\usepackage{color}
\usepackage{cleveref}

\usepackage{appendix}

\usepackage [english]{babel}
\usepackage [autostyle, english = american]{csquotes}
\MakeOuterQuote{"}
\usepackage[normalem]{ulem}

\makeatletter
\newcommand*{\rom}[1]{\expandafter\@slowromancap\romannumeral #1@}
\makeatother
\makeatletter
\def\p@subsection{}
\makeatother

\begin{document}
\title{Particle rearrangement and softening contributions 
to the nonlinear mechanical response of glasses}

\author{Meng Fan}
\affiliation{Department of Mechanical Engineering and Materials Science, Yale University, New Haven, Connecticut, 06520, USA}
\affiliation{Center for Research on Interface Structures and Phenomena, Yale University, New Haven, Connecticut, 06520, USA}
\author{Kai Zhang} 
\affiliation{Department of Chemical Engineering, Columbia University, New York, New York 10027, USA}
\author{Jan Schroers}
\affiliation{Department of Mechanical Engineering and Materials Science, Yale University, New Haven, Connecticut, 06520, USA}
\affiliation{Center for Research on Interface Structures and Phenomena, Yale University, New Haven, Connecticut, 06520, USA}
\author{Mark D. Shattuck}
\affiliation{Department of Physics and Benjamin Levich Institute, The City College of the City University of New York, New York, New York, 10031, USA}
\affiliation{Department of Mechanical Engineering and Materials Science, Yale University, New Haven, Connecticut, 06520, USA}
\author{Corey S. O'Hern}
\affiliation{Department of Mechanical Engineering and Materials Science, Yale University, New Haven, Connecticut, 06520, USA}
\affiliation{Center for Research on Interface Structures and Phenomena, Yale University, New Haven, Connecticut, 06520, USA}
\affiliation{Department of Physics, Yale University, New Haven, Connecticut, 06520, USA}
\affiliation{Department of Applied Physics, Yale University, New Haven, Connecticut, 06520, USA}

\date{\today}

\begin{abstract}
Amorphous materials such as metallic, polymeric, and colloidal
glasses, exhibit complex preparation-dependent mechanical response to
applied shear. In particular, glassy solids yield, with a mechanical
response that transitions from elastic to plastic, with increasing
shear strain. We perform numerical simulations to investigate the
mechanical response of binary Lennard-Jones glasses undergoing
athermal, quasistatic pure shear as a function of the cooling rate $R$
used to prepare them.  The ensemble-averaged stress versus strain
curve $\langle\sigma(\gamma)\rangle$ resembles the spatial average in
the large size limit, which appears smooth and displays a putative
elastic regime at small strains, a yielding-related peak in stress at
intermediate strain, and a plastic flow regime at large strains.  In
contrast, for each glass configuration in the ensemble, the
stress-strain curve $\sigma(\gamma)$ consists of many short nearly
linear segments that are punctuated by particle-rearrangement-induced
rapid stress drops.  To explain the nonlinearity of
$\langle\sigma(\gamma)\rangle$, we quantify the shape of the small
stress-strain segments and the frequency and size of the stress drops
in each glass configuration.  We decompose the stress loss ({\it
  i.e.}, the deviation in the slope of $\langle\sigma(\gamma)\rangle$
from that at $\langle\sigma(0)\rangle$) into the loss from particle
rearrangements and the loss from softening ({\it i.e.}, the reduction
of the slopes of the linear segments in $\sigma(\gamma)$), and then
compare the two contributions as a function of $R$ and $\gamma$.  For
the current studies, the rearrangement-induced stress loss is larger
than the softening-induced stress loss, however, softening stress
losses increase with decreasing cooling rate. We also characterize the
structure of the potential energy landscape along the strain direction
for glasses prepared with different $R$, and observe a dramatic change
of the properties of the landscape near the yielding transition.  We
then show that the rearrangement-induced energy loss per strain can
serve as an order parameter for the yielding transition, which
sharpens for slow cooling rates and in the large system-size limit.

\end{abstract}


\maketitle

\section{Introduction} 
\label{sec:intro}

\begin{figure}
\begin{center}
\includegraphics[width=0.9\columnwidth]{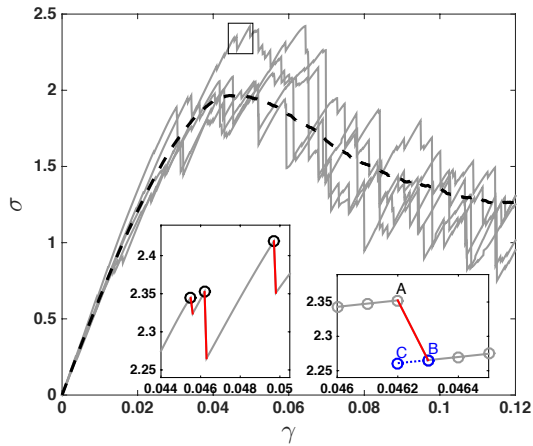}
\caption{The von Mises stress $\sigma$ versus strain $\gamma$ for five 
glass samples (dark gray curves) with $N=2000$ particles prepared at 
cooling rate $R=10^{-4}$ 
undergoing AQS pure shear. The ensemble-averaged 
stress $\langle \sigma \rangle$ over $500$ independent samples
is shown as the black dashed curve. The left inset provides a close-up of the 
stress in the strain interval $0.0440$ to $0.0505$, highlighting three 
of the stress drops with open circles at the start of the drop and red 
lines following the drop. The right inset is a close-up of the middle 
stress drop in the left inset. The gray solid line indicates forward 
shear increments of $\delta\gamma = 10^{-4}$, {\it i.e.} from point A at 
$\gamma=0.0462$ to point B at $\gamma + \delta\gamma=0.0463$. The blue dotted 
line indicates backward shear from point B at $\gamma+\delta\gamma=0.0463$ to 
point C at $\gamma=0.0462$. The magnitude of the shear stress difference 
for this rearrangement is $|\Delta \sigma(\gamma)|=|\sigma_A-\sigma_C|$ at $\gamma=0.0462$.}
\label{fig:intro}
\end{center}
\end{figure}

Glass formation occurs in many different materials, spanning an
enormous range of length scales, including atomic alloys, organic
compounds, ceramics, and dense colloidal
suspensions~\cite{ediger1996supercooled,berthier2011theoretical}.  In
particular, metallic glasses have received significant attention
recently for their promise in technological
applications~\cite{johnson1999bulk,inoue2007new,kanik2014high,kanik2015metallic}
that utilize their unique combination of properties ({\it e.g.} high
strength and elasticity) and
processability~\cite{ashby2006metallic,johnson2005universal,schroers2013bulk}.

Glasses are often generated by cooling a system in the liquid state
sufficiently rapidly such that crystallization is avoided and the
system remains disordered at low
temperatures~\cite{debenedetti2001supercooled}. The mechanical
response of glasses to applied stress is complex, including strain
hardening~\cite{deng2012atomistic}, plastic yielding~\cite
{karmakar2010statistical,hentschel2015stochastic,jaiswal2016mechanical,denisov2014sharp,
  kawasaki2016macroscopic,leishangthem2017yielding}, and brittle
failure~\cite{hays2000microstructure,lewandowski2005intrinsic,kumar2011unusual},
and the particular response that is observed for a given glass sample
depends on the protocol used to prepare and characterize it ({\it
  e.g.} its thermal history)~\cite{ketkaew2017fictive}. The cooling
rate determines the fictive temperature, at which the system falls out
of metastable equilibrium~\cite{moynihan1976dependence}.  The fictive
temperature significantly affects mechanical properties, such as the
ductility~\cite{fan2017effects,rycroft2012fracture,
  kumar2013critical,kumar2011unusual,li2013structural,nollmann2016impact},
shear band formation~\cite {zemp2015crystal}, quality factor of
vibrations~\cite{kanik2014high}, and the relation between stress
versus strain under quasistatic compression or
tension~\cite{utz2000atomistic,ashwin2013cooling}.

The fictive temperature defines the average energy of glasses in the
potential energy landscape
(PEL)~\cite{stillinger1995topographic,debenedetti2001supercooled},
which gives the potential energy of the system as a function of all of
the particle coordinates (and boundary conditions).  The PEL has been
used recently to describe the Gardner transition, the temperature
below which the separations between basins in the landscape becomes
fractal~\cite{charbonneau2014fractal,o2016signs}, super-Arrhenius
structural relaxation~\cite{yu2015strain}, as well as reversibility and
memory encoding during cyclic shear in
glasses~\cite{regev2015reversibility,fiocco2014encoding}.  Studies
have also quantified the width and depth of basins in the PEL using
thermal activation and saddle-point identification
methods~\cite{fan2014thermally,fan2015crossover}. This prior work
showed that the size of basins are smaller for more rapidly
cooled glasses, while asserting that the curvature of the basins is
insensitive to the cooling rate.  Other computational studies have
applied external shear to study the evolution of the system as the PEL
deforms with
strain~\cite{malandro1999relationships,lacks2004energy,maloney2006amorphous}, 
which strongly influences the mechanical response of glasses.  Instead 
of focusing on small strain intervals near mechanical instabilities, in this 
work, we will map out the full PEL in the strain direction. 

In contrast to crystalline materials, where the creation of and
interaction between topological defects controls the mechanical
response, it is more difficult to identify the structural defects that
control the mechanical response of glasses. In glasses, strong
non-affine motion in response to deformation is concentrated in ``shear
transformation zones''
(STZs)~\cite{argon1979plastic,falk1998dynamics,langer2008shear}.
Researchers have observed that particles occurring in STZs correlate
with those that possess low local yield stress~\cite{patinet2016connecting}
and participate in soft modes defined from the vibrational
density of states~\cite{widmer2008irreversible,ding2014soft}.  These 
prior studies have shown that with
increasing applied shear, the density of STZs increases, the elastic
regions decrease in size, STZs percolate, and plastic deformation
occurs~\cite{harmon2007anelastic}. Thus, rearrangements 
and non-affine motion in STZs also strongly impact mechanical response.    

In this article, we perform computer simulations of model structural
(binary Lennard-Jones) glasses undergoing athermal, quasistatic (AQS)
pure shear~\cite{maloney2006amorphous} to study their mechanical
response as a function of the cooling rate used to prepare them.  In
Fig.~\ref{fig:intro}, we show the shear stress versus strain during
AQS pure shear for five glass samples, all prepared at the same
cooling rate. While the ensemble-averaged shear stress versus strain
is smooth, the curve for each individual sample is
not~\cite{dubey2016elasticity}.  The left inset of
Fig.~\ref{fig:intro} shows that the shear stress versus strain curve
for a single configuration is composed of many nearly linear segments
punctuated by stress drops over narrow strain intervals.  Similar
behavior occurs for the total potential energy per particle $U$ (and
other quantities) versus strain, except in the case of $U$, the
segments are portions of parabolas. During the strain intervals with
continuous variation of $\sigma$ or $U$, the system remains in a
series of related minima in the PEL. At the strains corresponding to
the stress drops, the system becomes unstable, particles rearrange,
and the system evolves to a new lower minimum in the
PEL~\cite{malandro1999relationships,lacks2004energy,maloney2006amorphous}.

Thus, it is clear that the highly nonlinear shape of the
ensemble-averaged stress, potential energy, and other quantities
versus strain are determined by 1) the statistics of particle
rearrangements including both the frequency and size of rearrangements
and 2) changes in the form of the continuous regions of the piecewise
curves ({\it i.e.} softening or a decrease in the slopes of stress versus 
strain) between rearrangements.  There have been a number of
previous computational studies focusing on either particle
rearrangements~\cite
{fan2017effects,hentschel2015stochastic,lerner2009locality,karmakar2010statistical},
or softening of the shear
modulus~\cite{dubey2016elasticity,chikkadi2015spreading} of binary
Lennard Jones glasses under applied deformation.  In this article, we
will study {\it both} and compare the relative contributions from particle rearrangements
and softening in determining the ensemble-averaged nonlinear
mechanical response of sheared glasses as a function of the cooling
rate used to prepare them.
 
We seek to understand the highly nonlinear behavior of the
ensemble-averaged stress $\langle \sigma\rangle$ and potential energy
$\langle U\rangle$ versus strain (Fig.~\ref{fig:intro}).  Being able
to explain the {\it ensemble-averaged} mechanical response is
important for several reasons.  First, we will show below that the
system size dependence of ensemble-averaged quantities (like $\langle
\sigma\rangle$ and $\langle U\rangle$) is weak even for modest system
sizes. This result suggests that the ensemble average is
similar to the spatial average in the large system limit. Second, the
magnitude of the particle rearrangements decreases and the frequency
of particle rearrangements increases with increasing system
size~\cite{karmakar2010statistical}. Thus, it becomes increasingly
difficult to distinguish the continuous regions in the mechanical
response from drops due to particle rearrangements in the large system
limit.

As shown in Fig.~\ref{fig:intro}, as the applied shear strain
increases, the ensemble-averaged stress $\langle\sigma\rangle$ becomes
nonlinear in strain, the stress reaches a peak, and then decreases to a
plateau value in the large-strain limit. In the strain interval
between the strain at which the slope of the stress versus strain
curve begins to deviate significantly from that at zero strain and the
steady-state strain regime, yielding occurs and the system transitions
from an amorphous solid to a liquid-like state that can sample many
different configurations.  However, it is difficult to precisely
define the yielding transition from the smooth, ensemble-averaged
stress versus strain $\langle\sigma(\gamma)\rangle$~\cite{regev2013onset}.
There are many fundamental questions concerning yielding in glasses
since it involves a transition between two disordered states. For
example, does yielding represent a phase transition and, if so,
what is the appropriate order parameter that characterizes the
transition~\cite{hentschel2015stochastic}? 

Recent studies of the system-size scaling of particle rearrangement
statistics~\cite {hentschel2015stochastic,karmakar2010statistical},
configurational overlap between minima in the
PEL~\cite{jaiswal2016mechanical}, changes in symmetry of
nearest-neighbor structure~\cite{denisov2014sharp}, diffusivity of
rearranging particles~\cite {kawasaki2016macroscopic}, and onset of
irreversibility during cyclic
shear~\cite{regev2013onset,regev2015reversibility,leishangthem2017yielding},
have suggested that yielding in glassy materials can be described as a
non-equilibrium first-order phase transition.  For example, the
rearrangement frequency displays power-law scaling with system size,
with a scaling exponent that changes strongly as the strain approaches
the yield strain~\cite{hentschel2015stochastic}. Beyond the yield
strain, the exponent reaches a plateau value that is independent of
the cooling rate.  These studies have also shown that the yielding
transition becomes less sharp with increasing cooling
rate~\cite{hentschel2015stochastic}. However, much more work is needed
to fully understand the cooling-rate dependence of the rearrangement-
and softening-induced losses near yielding.

We present several key results in this study. First, we show that the
loss in stress from rearrangements is dominant over the
softening-induced stress loss, both of which have different cooling
rate and strain dependence.  Second, we quantify the rearrangement-
and softening-induced potential energy loss as a function of cooling
rate and strain. We measure the geometric features of the basins in
the PEL along the strain direction and find that the features of the
PEL change dramatically near yielding. Third, we propose additional
order parameters for the yielding transition based on the stress or
energy loss per strain from rearrangements and softening. The stress
(or energy) loss per strain increases rapidly near yielding and
increasing the system size leads to a sharper transition. Finally, we
calculate the distribution of energy drops from rearrangements as a
function of strain for different cooling rates. We find that the
distribution of energy drops is exponential with an energy scale that
also changes dramatically near yielding.

The remainder of the article is organized as follows.  In
Sec.~\ref{sec:methods}, we describe the computer simulations used to
prepare and shear the glasses at zero temperature, the physical
quantities that will be measured during the applied shear, and the
method employed to decompose the shear stress and potential energy
losses into contributions from rearrangements and from softening.
Sec.~\ref{sec:results} presents the results from the stress and energy
loss decompositions. We also characterize the geometric features of
PEL basins along the strain direction. In addition, we identify
quantities that change significantly with strain near yielding and
assess system-size effects. In Sec.~\ref{sec:conclusions}, we present
our conclusions and describe promising future research directions
concerning sheared glasses.

\section{methods} 
\label{sec:methods}

Our computational studies focus on model binary Lennard-Jones
mixtures, which have been shown to be good glass-formers. The computer
simulations are carried out in three stages: 1) Initialization of the
liquid state; 2) Cooling the liquid state to a zero-temperature glass
at a given rate and fixed low pressure; and 3) Application of
AQS pure shear deformation at fixed low pressure.
 
We first perform molecular dynamics (MD) simulations of binary
Lennard-Jones liquids in three dimensions under periodic boundary
conditions with constant particle number $N$ and pressure $P$.  We
consider $80\%$ large (type $A$) and $20\%$ small (type $B$) spherical
particles by number (both with mass $m$) in a cubic box with volume
$V$.  The particles interact pairwise via the shifted-force version of
the Lennard-Jones potential, $u(r_{ij}) = 4
\epsilon_{ij}[(\sigma_{ij}/r_{ij})^{12}-(\sigma_{ij}/r_{ij})^6]$ with
a cutoff distance $r_c =2.5 \sigma_{ij}$, where $r_{ij}$ is the
separation between particles $i$ and $j$.  The energy and length
parameters follow the Kob-Andersen mixing rules~\cite{kob1995testing}:
$\epsilon_{AA}=1.0$, $\epsilon_{BB}=0.5$, $\epsilon_{AB}=1.5$,
$\sigma_{AA}=1.0$, $\sigma_{BB}=0.88$, and $\sigma_{AB}=0.8$.  Length,
energy, temperature, pressure, and time scales are expressed in units
of $\sigma_{AA}$, $\epsilon_{AA}$, $\epsilon_{AA}/k_B$,
$\epsilon_{AA}/\sigma_{AA}^3$, and $\sigma_{AA}
\sqrt{m/\epsilon_{AA}}$, respectively, where $k_B$ is Boltzmann's
constant~\cite{allen1989computer}. We considered systems with $N=250$,
$500$, $1000$, $2000$, and $4000$ particles to study system-size effects.

To set the temperature and pressure, we incorporate a Nos\'{e}-Hoover
thermostat and barostat and integrate the equations of motion using a
second-order simplectic integration
scheme~\cite{plimpton1995fast,tuckerman2006liouville} with time step
$\Delta t = 10^{-3}$. We first equilibrate systems in the liquid
regime at constant temperature $T_0 = 0.6$ and pressure $P=0.025$ with
randomized initial particle positions and velocities. We then cool the
systems into a glassy state at zero temperature using a linear cooling
ramp, $T(t) = T_0-R t$, over a range of cooling rates from $R =
10^{-1}$ to $10^{-6}$, all of which are above the critical cooling
rate to ensure all zero-temperature samples are disordered. For each
cooling rate, we consider at least $N_c = 500$ configurations with
random initial conditions.

After generating each zero-temperature glass, we apply AQS pure shear
(AQS) at fixed pressure. For each strain step, we expand the box
length and move all particles affinely in the $x$-direction by a small
strain increment $\delta\gamma_x=\delta\gamma=10^{-4}$ and compress
the box length and move all particles affinely in the $y$-direction by
the same strain increment $\delta\gamma_y=-\delta\gamma$. Following
each strain step, we minimize the total enthalpy $H={\cal U} + PV$ at
fixed pressure $P=10^{-8}$, where ${\cal U}=\sum_{i>j} u(r_{ij})$ is
the total potential energy. We successively apply affine shear
increments $\delta\gamma$ followed by potential energy minimization to
a total strain $\gamma$.  Additional details concerning the AQS pure
shear algorithm can be found in our previous
studies~\cite{fan2017effects}.

We monitor the total 
potential energy per particle
$U(\gamma)={\cal U}(\gamma)/N$ and von Mises stress $\sigma(\gamma)$
as a function of strain $\gamma$ during the pure shear deformation.
The $3\times3$ stress tensor is given by
\begin{equation}
\label{eq:st}
{\Sigma}_{\mu \delta}=\frac{1}{V}\sum_{i>j}f_{ij\mu}r_{ij\delta},
\end{equation}
where $f_{ij\mu}$ is the $\mu=x,y,z$ component of the pairwise
force ${\vec f}_{ij}$ that particle $j$ exerts on particle $i$,
and $r_{ij\delta}$ is the $\delta=x,y,z$ component of the center-to-center
distance vector ${\vec r}_{ij}$ between particles $i$ and $j$.  The
von Mises stress $\sigma$ is the second
invariant of the stress tensor:
\begin{equation}
\label{eq:von}
\sigma=\sqrt{\frac{3}{2}{\rm Tr}(\pmb{\Sigma}+P\pmb{I})^2},
\end{equation}
where $\pmb{I}$ is the identity tensor and $P=-{\rm Tr}\pmb{\Sigma}/3$ is
the pressure~\cite{utz2000atomistic}.  We subtract the residual stress tensor
$\pmb{\Sigma}(\gamma=0)$ from $\pmb{\Sigma}(\gamma)$ so that
the von Mises stress $\sigma(\gamma)$ is initialized to zero at $\gamma=0$. 

As described in Sec.~\ref{sec:intro}, nonlinearity in
ensemble-averaged quantities, such as $\langle U(\gamma) \rangle$ and
$\langle \sigma(\gamma)\rangle$, is caused by particle rearrangements
and changes to the forms of the piecewise segments of $U(\gamma)$ and
$\sigma(\gamma)$ between rearrangements. We analyze the relative
contributions of the two effects by defining:
\begin{equation}
\label{sigtotal}
\sigma(\gamma)=\sigma_{\rm elastic}(\gamma) - \sigma_{\rm loss}(\gamma) - \sigma_{\rm loss}'(\gamma),
\end{equation}
where $\sigma_{\rm elastic}(\gamma)$ is the stress in the absence 
of losses from rearrangements and softening,
$\sigma_{\rm loss}(\gamma)$ is the loss in stress from 
particle rearrangements, and 
$\sigma_{\rm loss}'(\gamma)$ is the loss in stress from softening.  We 
define a similar expression for the total potential energy per particle: 
\begin{equation}
\label{utotal}
U(\gamma)=U_{\rm elastic}(\gamma) - U_{\rm loss}(\gamma) - U_{\rm loss}'(\gamma),
\end{equation}
where $U_{\rm elastic}(\gamma)$ is potential energy in the absence of
losses from rearrangements and softening, and $U_{\rm loss}(\gamma)$ and
$U_{\rm loss}'(\gamma)$ give the potential energy loss from
rearrangements and softening, respectively.

\begin{figure}
\begin{center}
\includegraphics[width=0.9\columnwidth]{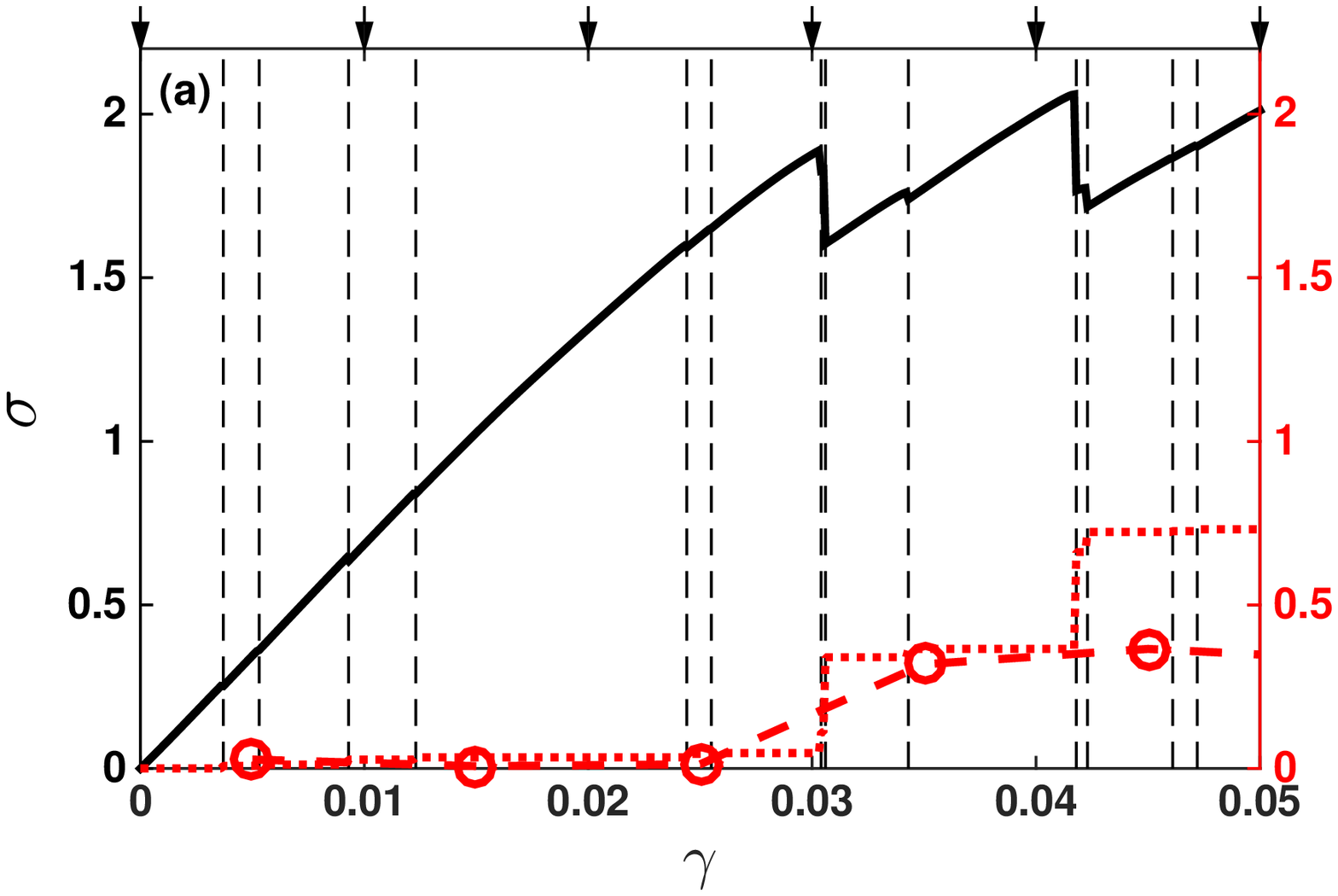}
\includegraphics[width=0.9\columnwidth]{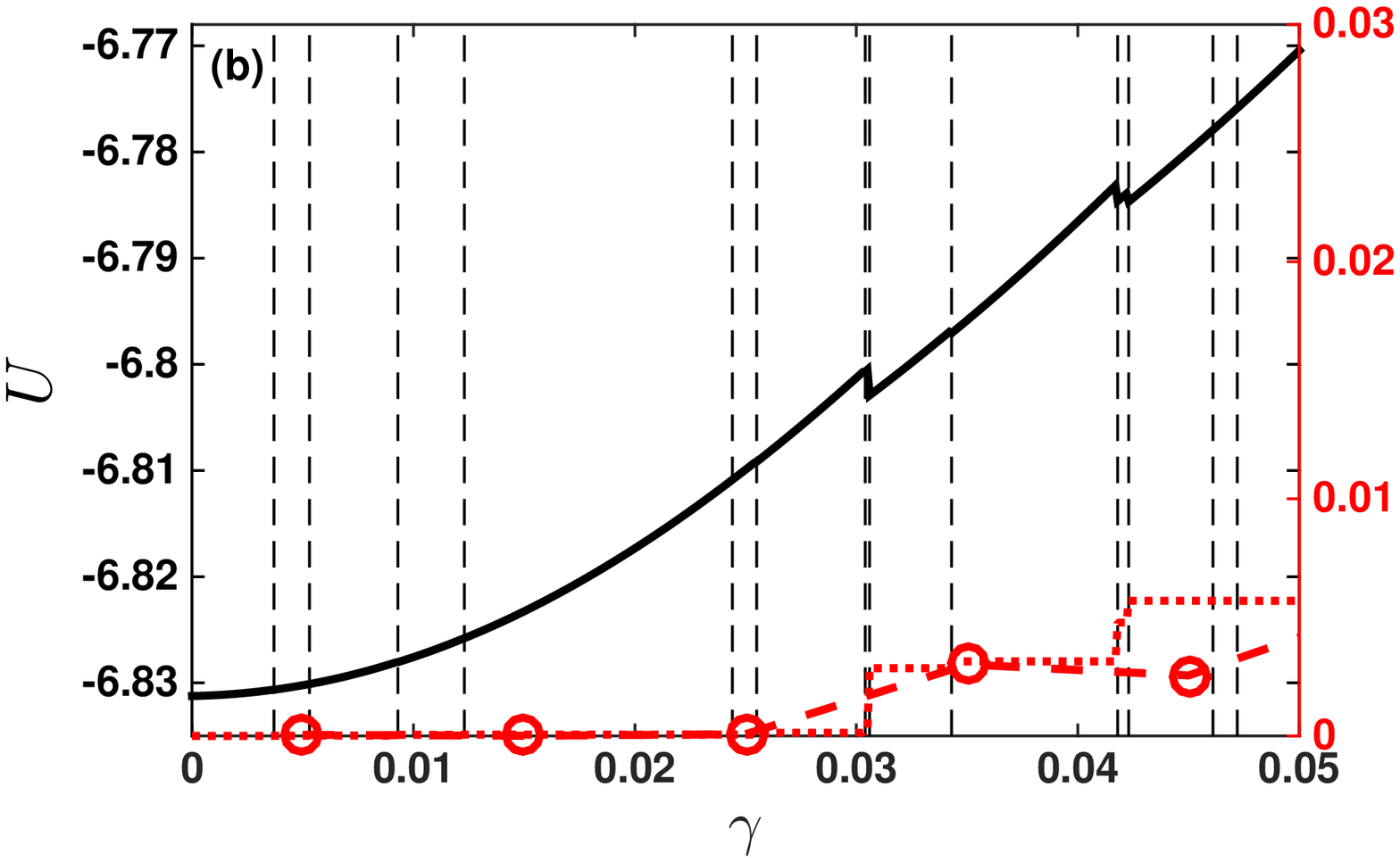}
\caption{Method to quantify the rearrangement-induced stress and energy losses. (a) The von Mises stress $\sigma$ and (b) total potential energy per
particle $U$ plotted versus strain $\gamma$ (solid black curves) for
a single glass configuration prepared at $R=10^{-4}$ undergoing
AQS pure shear. The vertical dashed lines indicate
the strains at which rearrangements occur. The rearrangement-induced
cumulative stress loss $\sigma_{\rm loss}$ (Eq.~\ref{sigloss}) and
potential energy loss $U_{\rm loss}$ (Eq.~\ref{uloss}) are shown by
the red dotted curves, with tick marks on the right vertical
axis. The binned rearrangement-induced stress and energy loss per (1\%)
strain, $d\sigma_{\rm loss}/d\gamma$ (Eq.~\ref{siglossr}) and
$dU_{\rm loss}/d\gamma$ (Eq.~\ref{ulossr}), are indicated by red
circles and dashed lines with tick marks on the right vertical
axis. The strain dependence has been binned with width
$d\gamma=0.01$; the edges of each bin are indicated by black arrows
in panel (a).}
\label{fig:methodrearrange}
\end{center}
\end{figure}

In previous simulation studies of AQS pure
shear~\cite{fan2017effects}, we developed a method to
unambiguously determine whether a particle rearrangement event occurs
during the strain interval $\gamma$ to $\gamma+\delta\gamma$ with an
accuracy on the order of numerical precision.  We denote the total
number of rearrangements in the strain interval $0$ to $\gamma$ as
$N_r(\gamma)$.  We calculate the cumulative rearrangement-induced
stress and energy loss after the $N_r(\gamma)$ rearrangements
in the strain interval $0$ to $\gamma$ as:
\begin{equation}
\label{sigloss}
\sigma_{\rm loss}(\gamma) = \sum_{i=1}^{N_r(\gamma)} |\Delta \sigma(\gamma_i)|
\end{equation}
\begin{equation}
\label{uloss}
U_{\rm loss}(\gamma) = \sum_{i=1}^{N_r(\gamma)} |\Delta U(\gamma_i)|,
\end{equation}
where $\gamma_i$ indicates the strains at which
rearrangements occur and $\Delta\sigma(\gamma_i)$ and $\Delta U(\gamma_i)$
are the stress and potential energy drops at each rearrangement, respectively.
(See the right inset of Fig.~\ref{fig:intro}.) 
We also measure the rearrangement-induced
stress and potential energy losses per strain:
\begin{equation}
\label{siglossr}
\left[\frac{d \sigma_{\rm loss}}{d\gamma}\right]\left(\gamma\right) = \frac{\sigma_{\rm loss}(\gamma+d \gamma)-\sigma_{\rm loss}(\gamma)}{d \gamma}
\end{equation}
\begin{equation}
\label{ulossr}
\left[\frac{d U_{\rm loss}}{d \gamma}\right]\left(\gamma\right) = \frac{U_{\rm loss}(\gamma+d \gamma)-U_{\rm loss}(\gamma)}{d \gamma},
\end{equation}
using bins of width $d\gamma = 10^{-2}$.  The stress and potential
energy losses from rearrangements ($\sigma_{\rm loss}$ and $U_{\rm
  loss}$), as well as the corresponding losses per strain
($d\sigma_{\rm loss}/d\gamma$ and $dU_{\rm loss}/d\gamma$), are shown
in Fig.~\ref{fig:methodrearrange} for a single configuration prepared
at $R=10^{-4}$.

\begin{figure}
\begin{center}
\includegraphics[width=0.9\columnwidth]{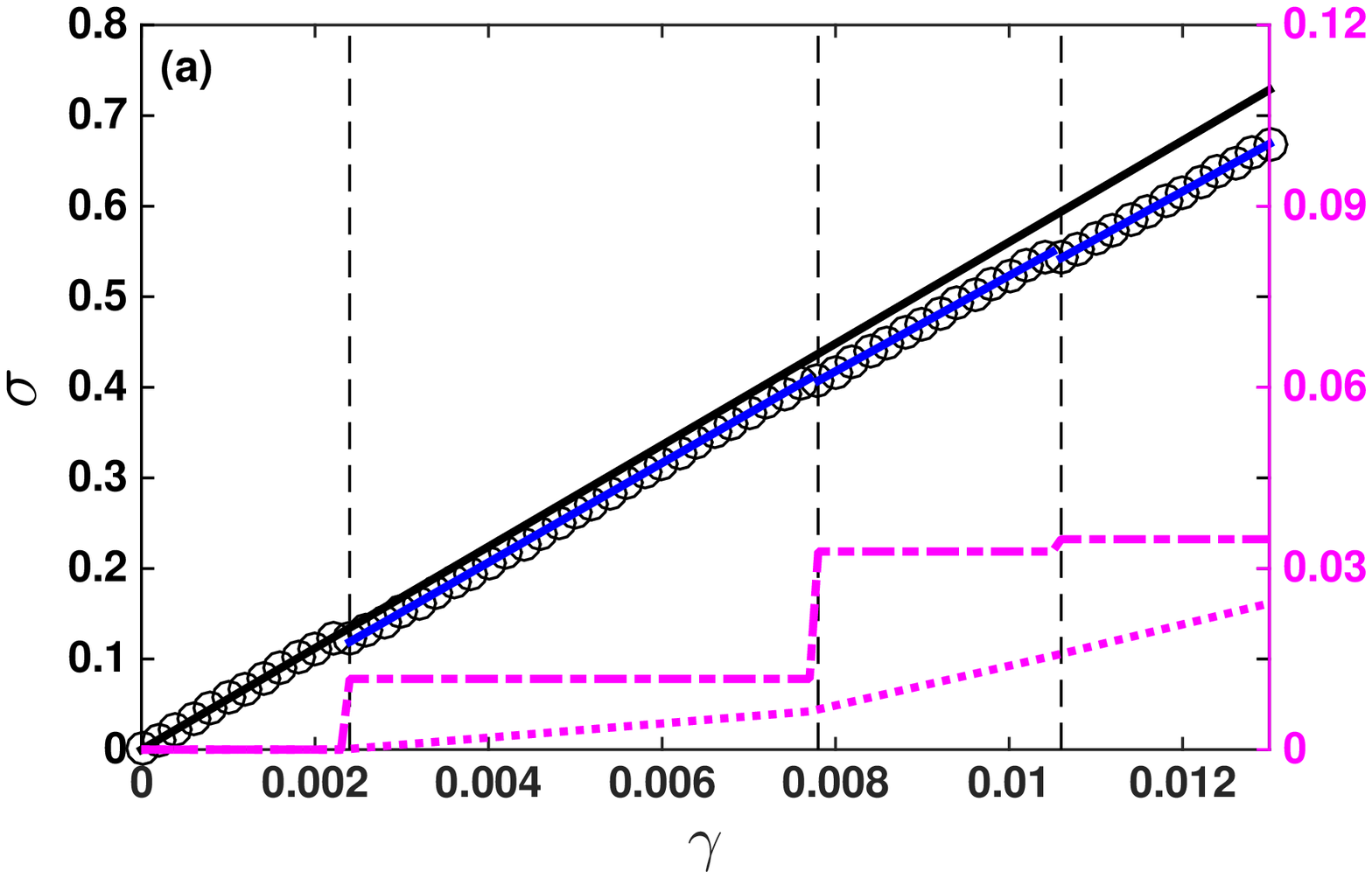}
\includegraphics[width=0.9\columnwidth]{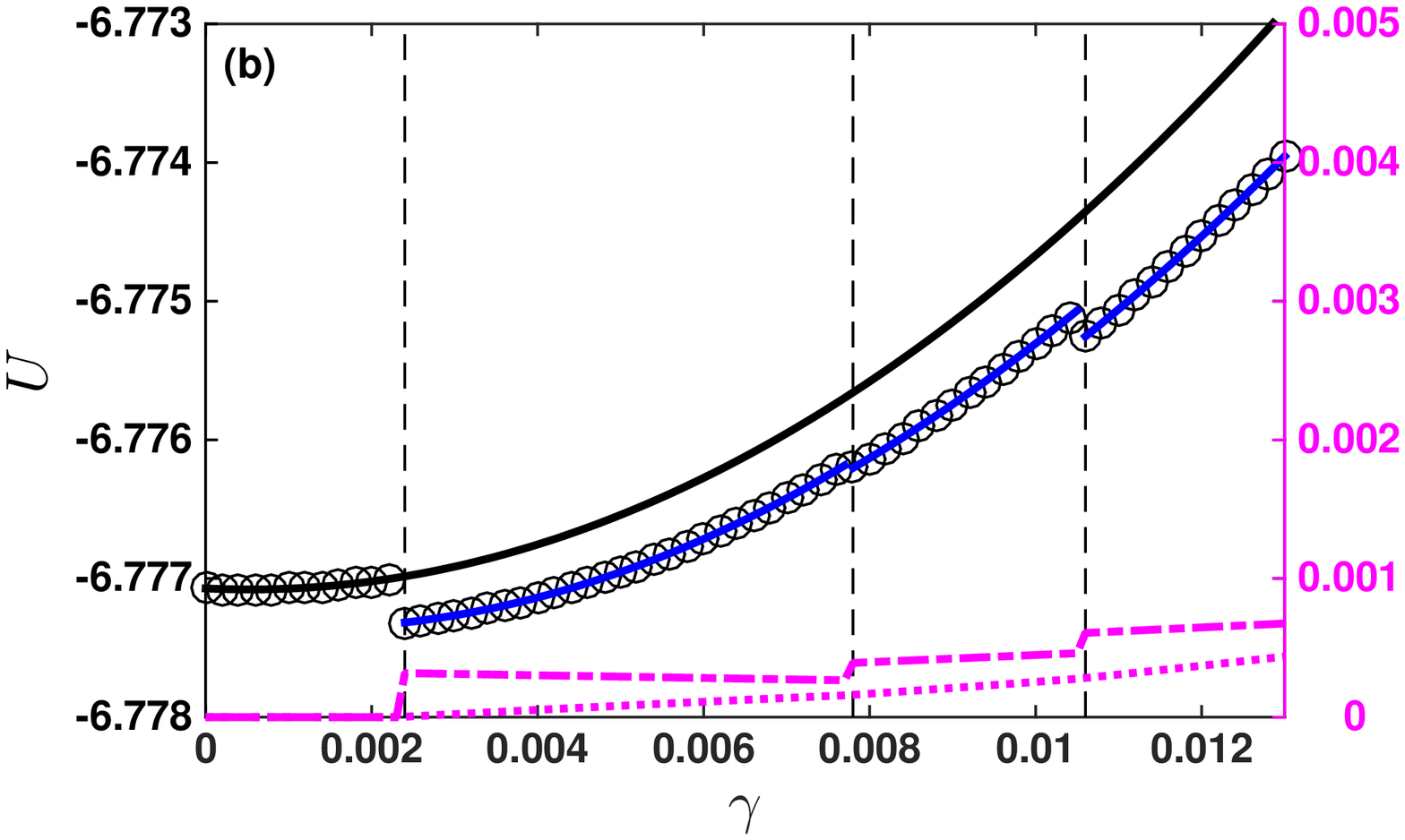}
\caption{Method to quantify the softening-induced stress and energy losses. (a) The von Mises stress $\sigma$ and (b) total potential energy per
particle $U$ plotted versus strain $\gamma$ (open black circles) for
a single glass configuration prepared at $R=10^{-2}$ undergoing
AQS pure shear.  Three rearrangements (indicated by
dashed vertical lines) occur in the strain interval $0 < \gamma <
0.014$. In (a), the regions of $\sigma(\gamma)$ between each
rearrangement are nearly linear. Best-fit blue lines are shown for
each segment. The solid black line has slope $G_0$ representing the 
slope of $\sigma(\gamma)$ near
$\gamma=0$. The stress loss per (1\%) strain ($d\sigma_{\rm
loss}'/d\gamma$) arising from changes in the slope of the line
segments (magenta dot-dashed curve with tick labels on the right
vertical axis) and the corresponding cumulative stress loss
$\sigma_{\rm loss}'$ (magenta dotted curve with tick labels on the
right vertical axis) are also shown.  In (b), the regions of
$U(\gamma)$ between each rearrangement are parabolic.  The best-fit
parabolas (blue curves) for each strain interval are shown. The solid
black curve is the best-fit parabola $U_0(\gamma)$ for the potential
energy near $\gamma=0$.  The potential energy loss per (1\%) strain
$dU_{\rm loss}'/d\gamma$ arising from changes in the local slope of
$U(\gamma)$ (magenta dot-dashed curve with tick marks on the right
vertical axis) and the cumulative potential energy loss $U_{\rm
loss}'$ (magenta dotted curve with tick marks on the right
vertical axis) are also shown.}
\label{fig:methodsoft}
\end{center}
\end{figure}


In Fig.~\ref{fig:methodsoft}, we illustrate how we quantify the effect
of softening on $\sigma(\gamma)$ and $U(\gamma)$.  In
panel (a), we show $\sigma(\gamma)$ for a single configuration as open
circles over a small strain interval. $\sigma(\gamma)$ is nearly
linear in regions of strain between the three stress drops that 
are indicated by dashed vertical lines. We define
the stress loss per strain from softening as:
\begin{equation}
\label{slr}
\left[\frac{d \sigma_{\rm loss}'}{d\gamma}\right] (\gamma) = G_{0}-G(\gamma),
\end{equation}
where $G_{0}$ is the slope of $\sigma(\gamma)$ in the $\gamma=0$ limit
(solid black line in Fig.~\ref{fig:methodsoft} (a)) and
$G(\gamma)$ is the slope of $\sigma(\gamma)$ at strain $\gamma$ (solid
blue lines in Fig.~\ref{fig:methodsoft} (a)).  $d\sigma_{\rm
  loss}'/d\gamma$ is a constant for each piecewise linear
stress-strain segment and is discontinuous at rearrangements. We also
measure the cumulative softening-induced stress loss for strain
up to $\gamma$ by integrating the corresponding stress loss per strain:
\begin{equation}
\label{sl}
\sigma_{\rm loss}'(\gamma)=\int_0^{\gamma} (d\sigma_{\rm loss}'/d\gamma') d\gamma',
\end{equation}
which is continuous, but the slope of the curve 
changes discontinuously at each rearrangement. 

To quantify the effect of softening on the potential energy versus strain 
$U(\gamma)$ (Fig.~\ref{fig:methodsoft} (b)), we find the best-fit parabola
for each piecewise elastic segment between rearrangement events using
\begin{equation}
\label{parafit}
U(\gamma)= \frac{A}{2} \gamma^2 + B \gamma + C,
\end{equation}
where $A$, $B$, and $C$ are coefficients that determine the concavity
and location of the parabola.  We define the potential energy loss per
strain from softening as $dU_{\rm loss}'/d\gamma = k_{0}-k(\gamma)$,
where $k_{0}$ and $k$ are the local slopes of $U(\gamma)$ at
strains $0$ and $\gamma$, respectively. Using Eq.~\ref{parafit}, we
define the softening-induced potential energy loss per strain as:
\begin{equation}
\label{ulr}
\left[\frac{dU_{\rm loss}'}{d\gamma}\right] (\gamma) = (A_{0}-A(\gamma))\gamma + B_{0}-B(\gamma),
\end{equation}
where the coefficients $A_{0}$ and $B_{0}$ are measured at $\gamma=0$.
In contrast to $d\sigma_{\rm loss}'/d\gamma$, which is constant,
$dU_{\rm loss}'/d\gamma$ depends linearly on $\gamma$ for each
inter-rearrangement segment. The cumulative softening-induced potential energy
loss can be calculated by integrating $dU'_{\rm loss}/d\gamma$ over 
a given strain interval:
\begin{equation}
\label{ul}
U_{\rm loss}'(\gamma)=\int_0^{\gamma} (dU_{\rm loss}'/d\gamma') d\gamma'.
\end{equation}
$U_{\rm loss}'(\gamma)$ is piecewise quadratic, whereas $\sigma_{\rm loss}'(\gamma)$ is piecewise linear.

\section{Results} 
\label{sec:results}

The discussion of the results is organized into three
subsections. First, in Sec.~\ref{sec:results_stress}, we illustrate
the effects of rearrangements and softening on the ensemble-averaged
stress versus strain curve as a function of the cooling rate. In
particular, we compare the relative contributions of rearrangements
and softening to the nonlinear mechanical response. In
Sec.~\ref{sec:results_energy}, we identify the distinct contributions
of rearrangements and softening to the loss in 
potential energy as a function of strain.  In addition, we study the
properties of the parabolic segments of $U(\gamma)$ between
rearrangements to characterize the width and height of
basins in the PEL near the yielding
transition.  In Sec.~\ref{sec:results_Neffect}, we investigate the
system-size scaling exponents for the size and frequency of
rearrangements and the distribution of energy drops from
rearrangements near the yielding transition. In addition, we study 
the stress and energy losses from rearrangements and softening 
as a function of system size.

\subsection{Stress losses from rearrangements and softening}
\label{sec:results_stress}

\begin{figure}
\begin{center}
\includegraphics[width=1.0\columnwidth]{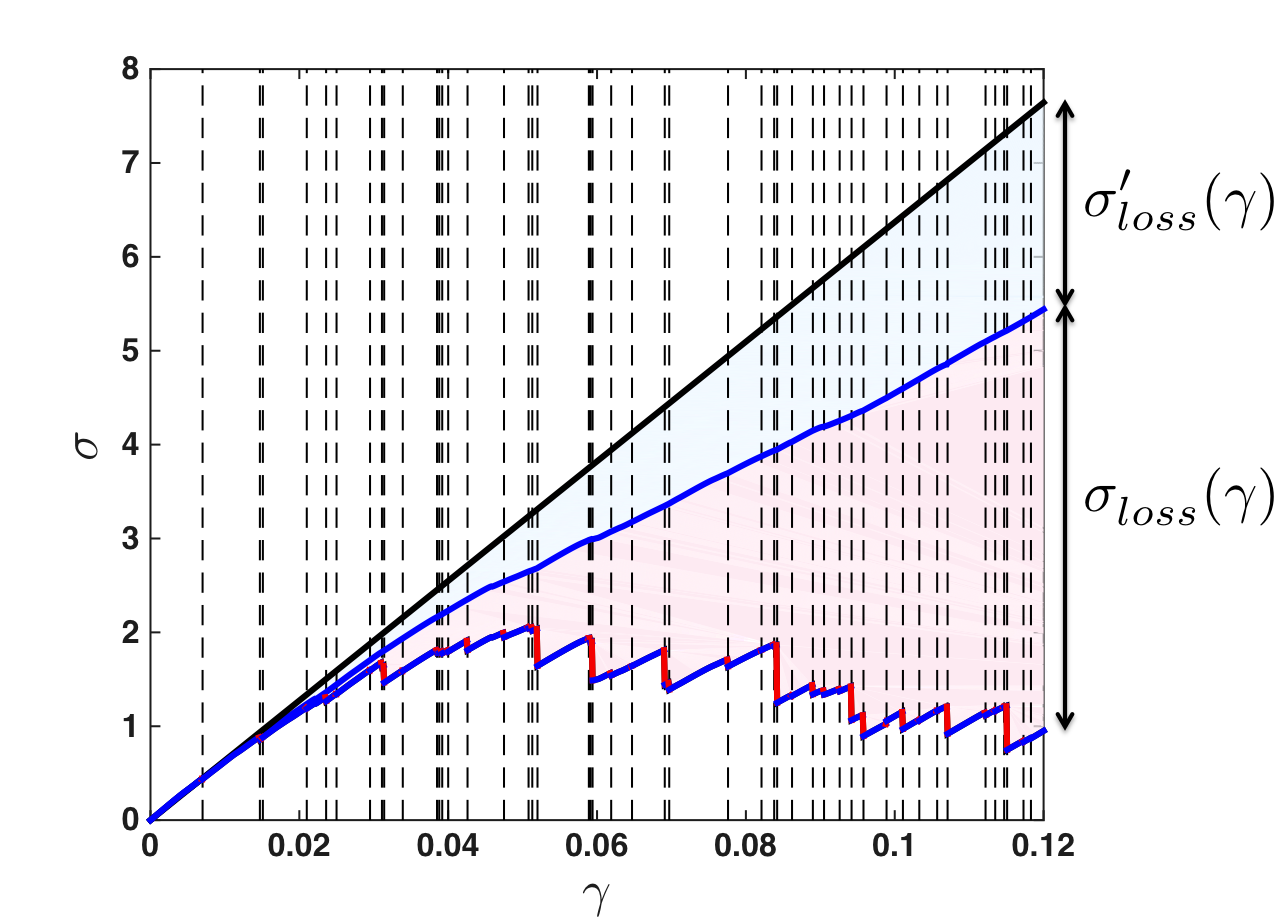}
\caption{Rearrangement and softening effects on the von Mises stress 
$\sigma$ for a single glass configuration
(with $N=2000$ and $R=10^{-4}$) undergoing AQS pure
shear. $\sigma(\gamma)$ (bottom
curve) has nearly linear continuous segments (blue lines)
punctuated by rapid stress drops caused by particle rearrangements (red
lines).  The strains at which the rearrangements occur are indicated 
by dashed vertical lines. The middle blue solid curve is obtained 
by connecting the continuous segments of $\sigma(\gamma)$ between 
rearrangements end to end. The stress $\sigma_{\rm elastic}$ in the 
absence of rearrangements and softening (top black line) is obtained 
from the slope of $\sigma(\gamma)$ in the $\gamma \rightarrow 0$ limit. 
The cumulative rearrangement-induced stress loss
$\sigma_{\rm loss}(\gamma)$ (Eq.~\ref{sigloss}) is defined as the 
width of the red-shaded region at each strain $\gamma$.  
The cumulative softening-induced stress loss
$\sigma_{\rm loss}'(\gamma)$ (Eq.~\ref{uloss}) is defined as the width of 
the blue-shaded area at each strain $\gamma$.}
\label{fig:example}
\end{center}
\end{figure}

In Fig.~\ref{fig:example}, we show the von Mises stress versus strain
$\sigma(\gamma)$ for a single glass configuration with $N=2000$
prepared at $R=10^{-4}$ undergoing AQS pure shear.
We identify the elastic contribution to the stress $\sigma_{\rm
  elastic}(\gamma)$ in the absence of rearrangements and softening,
the stress loss from rearrangements $\sigma_{\rm loss}(\gamma)$, and the stress loss from
softening $\sigma_{\rm loss}'(\gamma)$.  We find that both
$\sigma_{\rm loss}(\gamma)$ and $\sigma_{\rm loss}'(\gamma)$ increase
with strain. For most configurations including this one, the stress loss from
rearrangements is larger than that from softening, $\sigma_{\rm
  loss}(\gamma) > \sigma_{\rm loss}'(\gamma)$, and the difference grows 
with increasing strain.

\begin{figure}
\begin{center}
\includegraphics[width=1.0\columnwidth]{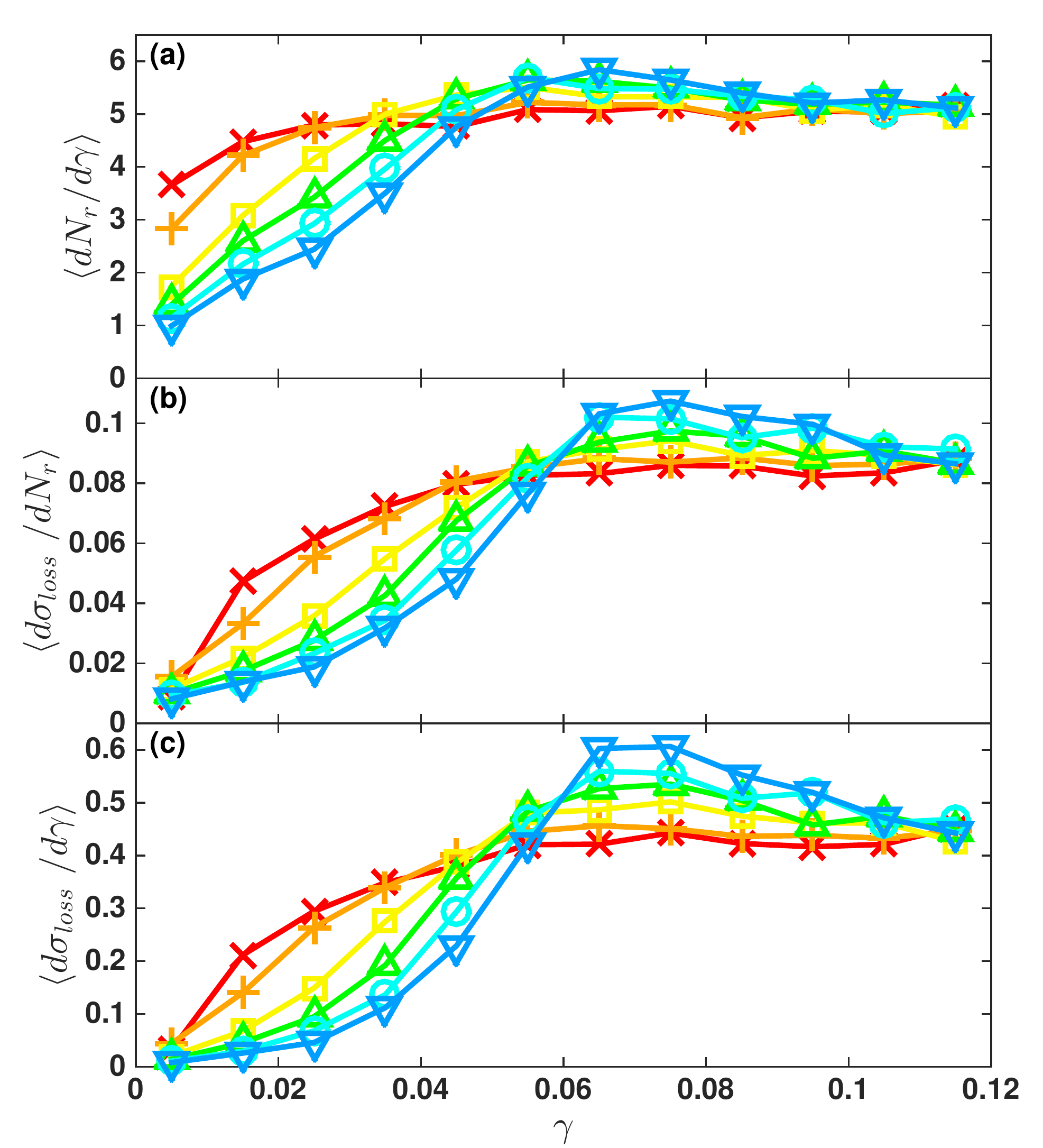}
\caption{Ensemble-averaged (a) rearrangement frequency $\langle dN_r/d\gamma \rangle$, (b)
stress loss per rearrangement $\langle d\sigma_{\rm loss}/dN_r \rangle$, and
(c) rearrangement-induced stress loss per (1\%) strain
$\langle d\sigma_{\rm loss}/d\gamma \rangle$ plotted versus
strain $\gamma$ for glasses undergoing AQS pure shear. 
The glasses were prepared using cooling rates $R=10^{-1}$ (crosses), $10^{-2}$ (plus signs), $10^{-3}$
(squares), $10^{-4}$ (upward triangles), $10^{-5}$ (circles), and $10^{-6}$ (downward triangles). 
All data is obtained by averaging over $500$ independent samples with $N=2000$.
}
\label{fig:RS}
\end{center}
\end{figure}

The stress loss per strain $d\sigma_{\rm loss}/d\gamma$ from
rearrangements (Eq.~\ref{siglossr}) can be decomposed as $d\sigma_{\rm
  loss}/d\gamma=(d\sigma_{\rm loss}/dN_r)(dN_r/d\gamma)$, where
$dN_r/d\gamma$ is the rearrangement frequency ({\it i.e.}, number of rearrangements per strain) and $d\sigma_{\rm
  loss}/dN_r$ is the rearrangement size ({\it i.e.}, stress loss per rearrangement).  In
Fig.~\ref{fig:RS}, we show the ensemble average of all three
quantities, $\langle dN_r/d\gamma \rangle$, $\langle d\sigma_{\rm loss}/dN_r \rangle$, and
$\langle d\sigma_{\rm loss}/d\gamma \rangle$, for glasses prepared over 
a range of cooling rates.  We find that all three increase at
small strains ($\gamma \lesssim 0.05$), plateau at large strains in
the steady state regime ($\gamma \gtrsim 0.1$), and form a peak in the
intermediate strain regime ($0.05 \lesssim \gamma \lesssim 0.1$).  The
peaks are more prominent for $\langle d\sigma_{\rm loss}/dN_r \rangle$ and
$\langle d\sigma_{\rm loss}/d\gamma \rangle$. At small strains, all three quantities
increase with cooling rate, indicating that rearrangements play a more
significant role in stress loss in more rapidly cooled glasses. In
contrast, at intermediate strains, all three quantities decrease with
increasing cooling rate.  In the large strain regime, none of the
quantities show cooling rate dependence.

\begin{figure}
\begin{center}
\includegraphics[width=1.0\columnwidth]{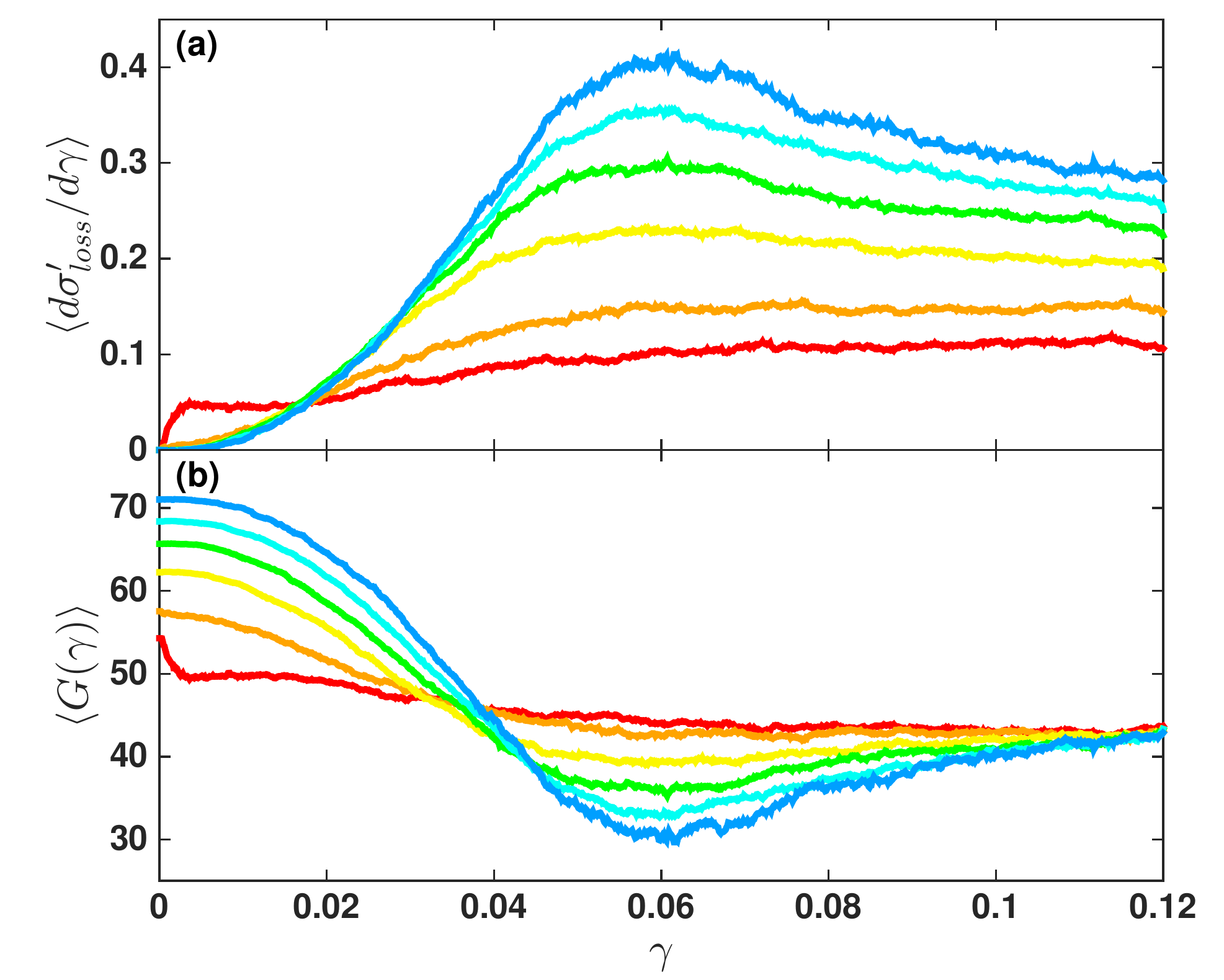}
\caption{Ensemble-averaged (a) softening-induced stress loss per (1\%) strain
$\langle d\sigma_{\rm loss}'/d\gamma \rangle$ and (b) local slope
of the continuous stress versus strain segments $\langle G(\gamma) \rangle$ plotted versus
strain $\gamma$ for cooling rates $R=10^{-1}$ (red), $10^{-2}$ (orange), $10^{-3}$
(yellow), $10^{-4}$ (green), $10^{-5}$ (cyan), and $10^{-6}$ (blue). 
All data is obtained by averaging over $500$ configurations with $N=2000$.
}
\label{fig:SS}
\end{center}
\end{figure}

In Fig.~\ref{fig:SS} (a), we show the ensemble-averaged
softening-induced stress loss per strain, $\langle d\sigma_{\rm
  loss}'/d\gamma \rangle$ (Eq.~\ref{slr}).  $\langle d\sigma_{\rm loss}'/d\gamma \rangle$
increases at small strains and plateaus at cooling rate-dependent values at large
strains. In the intermediate strain regime, $\langle d\sigma_{\rm
  loss}'/d\gamma \rangle$ is larger for more slowly cooled glasses with a
pronounced peak.  To gain insight into this behavior, we plot the
ensemble-averaged slope of the continuous stress versus strain
segments $\langle G(\gamma) \rangle$ in
Fig.~\ref{fig:SS} (b).  At $\gamma=0$, the shear modulus $\langle G(0)
\rangle$ depends on the degree of heterogeneity in the material and
thus increases with decreasing $R$~\cite{ashwin2013cooling}.  As
$\gamma$ increases, $\langle G(\gamma) \rangle$ decreases at small
strains and reaches a plateau value ($\approx 40$) at large strains
that is independent of cooling rate. In the intermediate strain regime, for
slowly cooled glasses, {\it e.g.} $R=10^{-6}$, $\langle
G(\gamma) \rangle$ first decreases near $\gamma \approx 0.04$
and reaches a minimum near $\gamma \approx 0.06$ corresponding to the 
peak in $\langle d\sigma_{\rm loss}'/d\gamma \rangle$. Thus, at small strains, the 
slowly cooled glasses are the most rigid, while at intermediate strains, they
are the least rigid. For rapidly cooled glasses, the non-monotonic 
behavior in strain is absent and the shear modulus $\langle G(\gamma)\rangle$ 
decreases continuously with strain until it plateaus in the large-strain limit. 

\begin{figure}
\begin{center}
\includegraphics[width=1.0\columnwidth]{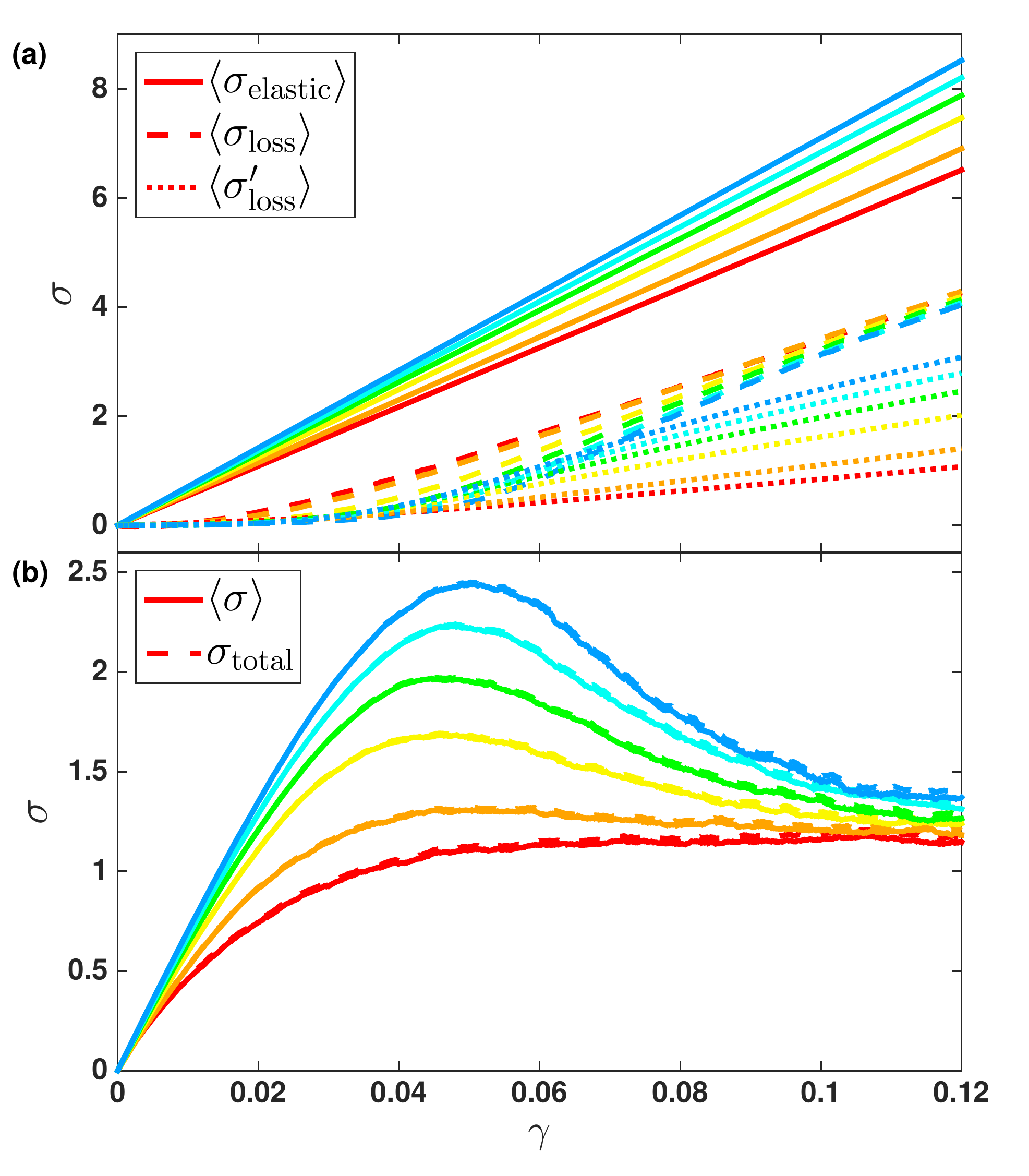}
\caption{Ensemble-averaged (a) elastic stress $\sigma_{\rm
elastic}(\gamma)$ (solid), rearrangement-induced stress loss $\sigma_{\rm
loss}(\gamma)$ (dashed), and softening-induced stress loss $\sigma_{\rm
loss}'(\gamma)$ (dotted) plotted versus strain $\gamma$.
(b) Ensemble-averaged stress $\langle \sigma(\gamma) \rangle$ and the 
stress $\sigma_{\rm total}$ obtained by combining the elastic stress, 
and the rearrangement- and 
softening-induced stress losses (Eq.~\ref{sigtotal}) plotted
versus $\gamma$.  In both (a) and (b), the system size $N=2000$, 
and six cooling rates, 
$R=10^{-1}$ (red), $10^{-2}$ (orange), $10^{-3}$
(yellow), $10^{-4}$ (green), $10^{-5}$ (cyan), and $10^{-6}$ (blue) 
are shown.}
\label{fig:stress}
\end{center}
\end{figure}

To compare the relative contributions of rearrangements and softening
on the stress loss, we integrate $d\sigma_{\rm loss}/d\gamma$ and
$d\sigma_{\rm loss}'/d\gamma$ over strain to obtain the cumulative
stress losses, $\sigma_{\rm loss}$ and $\sigma_{\rm loss}'$,
respectively.  In Fig.~\ref{fig:stress}, we show the ensemble average
of the four variables in Eq.~\ref{sigtotal}, as well as the direct
ensemble average $\langle \sigma(\gamma) \rangle$ of stress from
single glass configurations, for different cooling rates $R$. The
ensemble-averaged stress versus strain $\langle \sigma(\gamma)
\rangle$ and the combination of the terms in
Eq.~\ref{sigtotal}, $\sigma_{\rm total}(\gamma)$, agree quantitatively.

In general, $\langle \sigma_{\rm loss} \rangle > \langle \sigma_{\rm
  loss}' \rangle$, which means that stress losses from rearrangements
are larger than those from softening.  For the sake of discussion, we
divide the stress versus strain curve into three regions: the pre-peak
region ($\gamma \lesssim 0.04$), the peak region ($0.04 \lesssim
\gamma \lesssim 0.07$), and the post-peak region ($\gamma \gtrsim
0.07$).  In the pre-peak region, the stress loss from softening
$\langle \sigma_{\rm loss}' \rangle$ is extremely small, while the
stress loss from rearrangements $\langle \sigma_{\rm loss} \rangle$ is
nonzero and increases for more rapidly cooled glasses. Significant
stress loss from rearrangements in the pre-peak region for rapidly
cooled glasses explains the strongly nonlinear behavior of the stress
versus strain for large cooling rates $R$.  The ability of rapidly
cooled glasses to undergo rearrangements in the pre-peak strain region
is also correlated with enhanced ductility~\cite{fan2017effects}.  In
the peak region, the stress loss from softening $\langle \sigma_{\rm
  loss}'\rangle$ begins to grow and becomes comparable to the stress
loss from rearrangements $\langle \sigma_{\rm loss} \rangle$. However,
$\langle \sigma_{\rm loss} \rangle$ and $\langle \sigma_{\rm loss}'
\rangle$ display opposite cooling rate dependence. More rapidly cooled
glasses have larger stress loss from rearrangements and smaller stress
loss from softening in the peak region. In contrast, more slowly
cooled glasses possess smaller stress loss from rearrangements and
larger stress loss from softening.  In the post-peak region, both
$\langle \sigma_{\rm loss} \rangle$ and $\langle \sigma_{\rm loss}' \rangle$ increase linearly with
$\gamma$. At large strains, the stress loss from rearrangements
$\langle \sigma_{\rm loss}\rangle$ becomes cooling-rate independent. However, the
cooling rate dependence of the stress loss from softening $\langle \sigma_{\rm
  loss}' \rangle$ increases at large strains. In this region, $\langle \sigma_{\rm
  loss}' \rangle$ increases for more slowly cooled glasses, which gives rise
to the strong decay in $\langle \sigma(\gamma)\rangle$ at strains beyond the peak
stress. (See Fig.~\ref{fig:stress}.)

\subsection{Losses in potential energy and geometric features of basins in the energy landscape}
\label{sec:results_energy}

In this subsection, we will quantify the losses in the potential
energy $U$ from rearrangements and softening during AQS pure shear deformation. In addition, we will
characterize the geometric features of basins in the potential energy
landscape along the strain direction as a function of the cooling rate
$R$ used to prepare the glasses.

\begin{figure}
\begin{center}
\includegraphics[width=1.0\columnwidth]{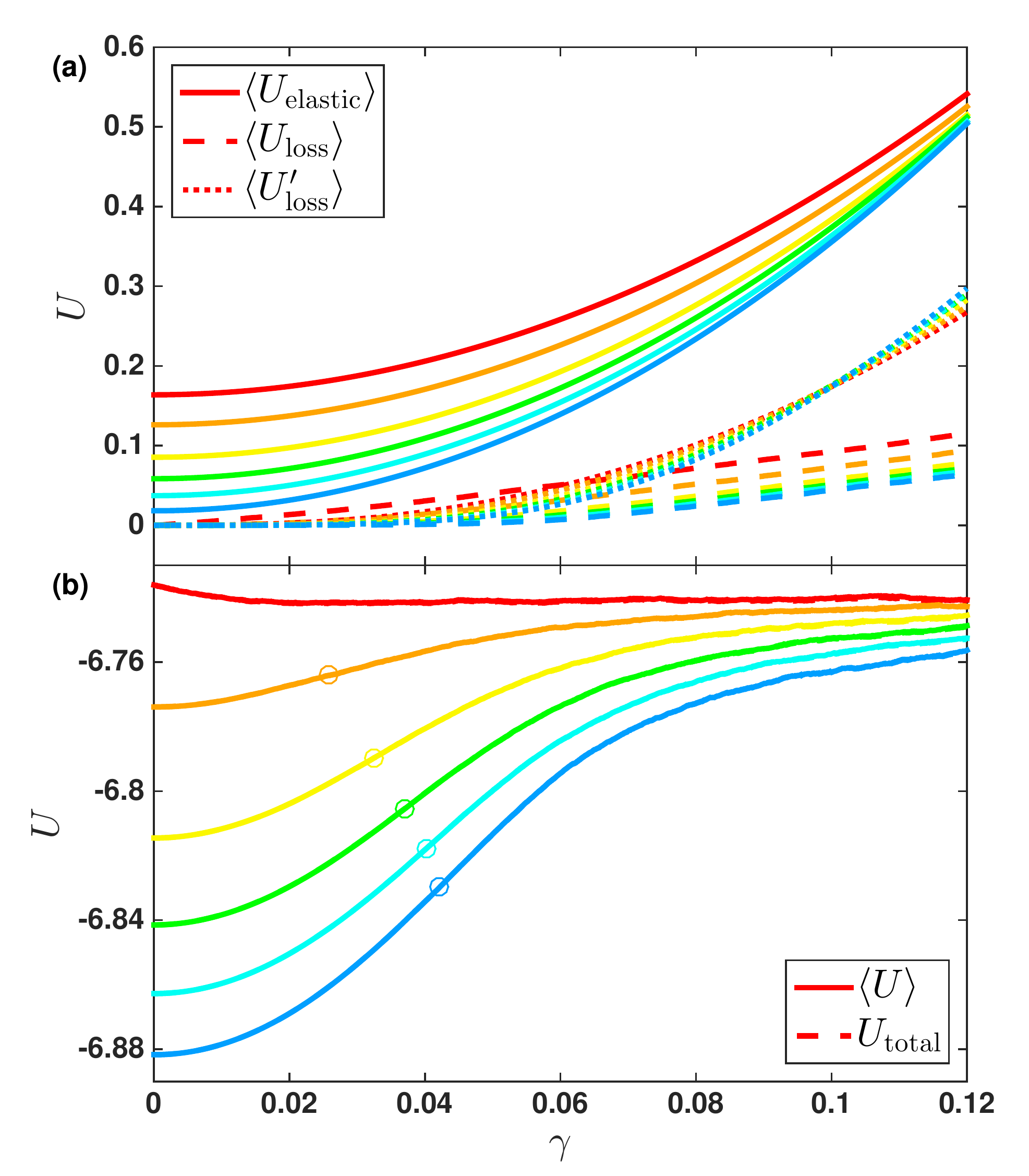}
\caption{Ensemble-averaged (a) elastic potential energy $\langle U_{\rm elastic}(\gamma) \rangle$,
rearrangement-induced potential energy loss $\langle U_{\rm loss}(\gamma)\rangle$, and softening-induced
potential energy loss $\langle U_{\rm loss}'(\gamma)\rangle$ plotted versus strain $\gamma$. (b) 
Ensemble-averaged potential energy 
$\langle U(\gamma) \rangle$ and the potential energy $U_{\rm total}(\gamma)$ 
obtained by combining the elastic energy, and the rearrangement- 
and softening-induced energy losses (Eq.~\ref{utotal}) plotted versus
$\gamma$ for cooling rates $R=10^{-1}$ (red), $10^{-2}$ (orange), $10^{-3}$
(yellow), $10^{-4}$ (green), $10^{-5}$ (cyan), and $10^{-6}$ (blue).
All data is obtained by averaging over $500$ samples with 
system size $N=2000$. The inflection points of $\langle U(\gamma) \rangle$ are indicated
by open circles.}
\label{fig:energy}
\end{center}
\end{figure}

In Fig.~\ref{fig:energy}, we show the ensemble-averaged potential
energy $\langle U(\gamma)\rangle$ and compare the potential energy
losses from rearrangements $\langle U_{\rm loss} \rangle$ and from
softening $\langle U_{\rm loss}'\rangle$ as a function of strain.  By
construction, the direct ensemble-averaged potential energy $\langle
U(\gamma)\rangle$ agrees quantitatively with the potential energy
$U_{\rm total}(\gamma)$ obtained by combining the terms $U_{\rm
  elastic}(\gamma)$, $U_{\rm loss}(\gamma)$, and $U_{\rm
  loss}'(\gamma)$ from Eq.~\ref{utotal}.  Near $\gamma=0$, $\langle
U(\gamma) \rangle$ is larger for more rapidly cooled glasses since
rapid cooling prevents the system from exploring configuration space
and finding lower energy
minima~\cite{debenedetti2001supercooled,utz2000atomistic}. At small
strains, $\langle U(\gamma) \rangle$ increases quadratically for all
cooling rates (except for $R=10^{-1}$) since the losses from
rearrangements and softening are small.

As $\gamma$ increases, the ensemble-averaged potential energy $\langle
U(\gamma)\rangle$ deviates from quadratic behavior due to increases in
losses from rearrangements $U_{\rm loss}(\gamma)$ and softening
$U_{\rm loss}'(\gamma)$. At large strains, $\langle U(\gamma)\rangle$
approaches a plateau value that is independent of the cooling rate
$R$~\cite{utz2000atomistic}.  As shown in Fig.~\ref{fig:energy} (a),
the potential energy loss from rearrangements $\langle U_{\rm loss}(\gamma)\rangle$
increases with cooling rate for all $\gamma$~\cite{fan2017effects}. In
contrast to the behavior for the stress losses from rearrangements
(Fig.~\ref{fig:stress} (a)), the potential energy loss from
rearrangements $\langle U_{\rm loss}(\gamma) \rangle$ is smaller than the potential
energy loss from softening $\langle U_{\rm loss}'(\gamma)\rangle$. In fact, at large
strains, $\langle U_{\rm loss}'(\gamma)\rangle$ grows more rapidly with strain than
$\langle U_{\rm loss}(\gamma) \rangle$.  The strain dependence of $\langle U_{\rm
  loss}'(\gamma)\rangle$ originates from the evolution with strain of the
geometric features of the PEL ({\it i.e.}, the
$\gamma$-dependence of the two terms, $(A_0-A(\gamma))\gamma$ and
$B_0-B(\gamma)$, in Eq.~\ref{ulr}), which will be discussed below.

\begin{figure}
\begin{center}
\includegraphics[width=0.9\columnwidth]{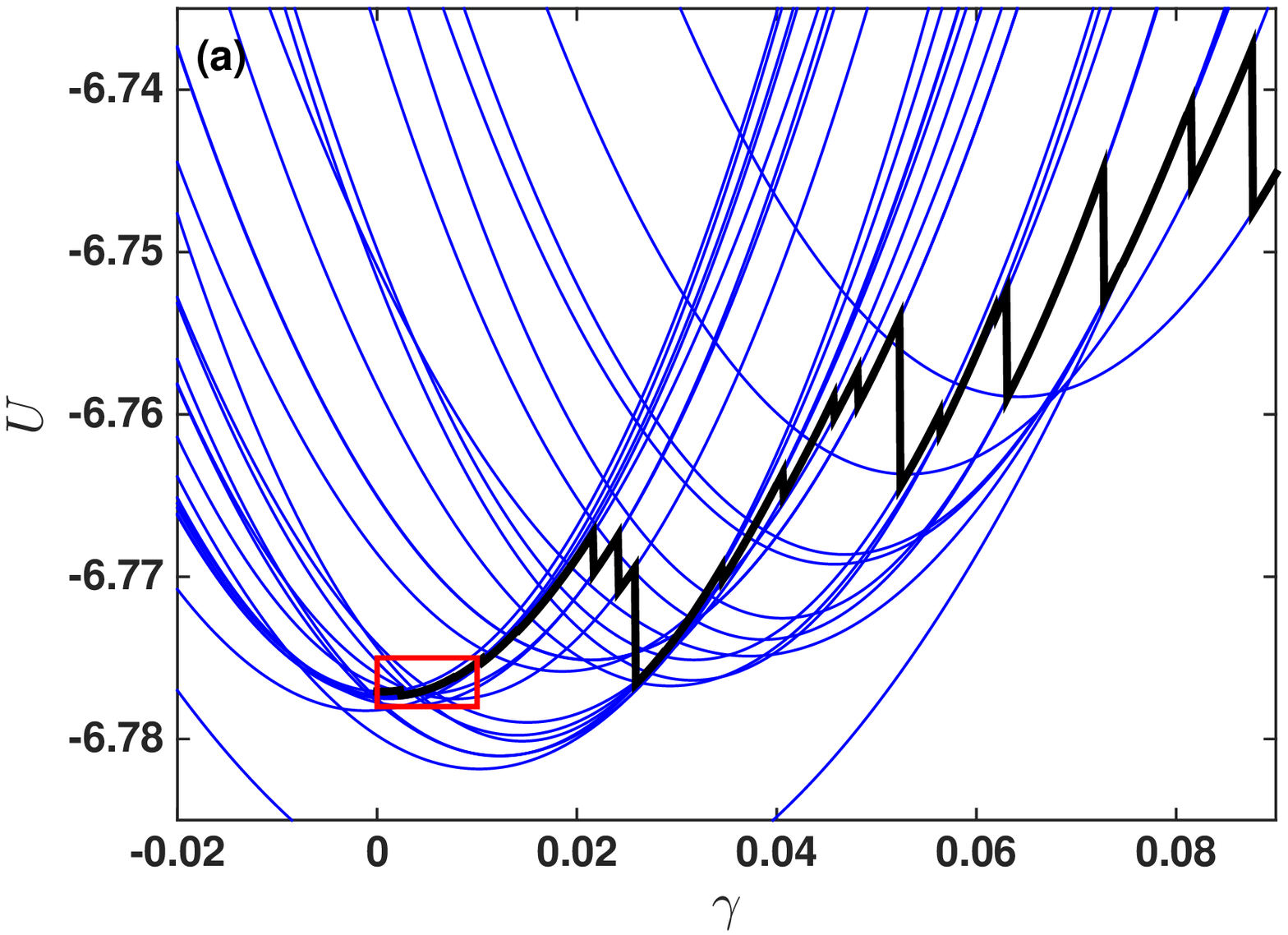}
\includegraphics[width=0.9\columnwidth]{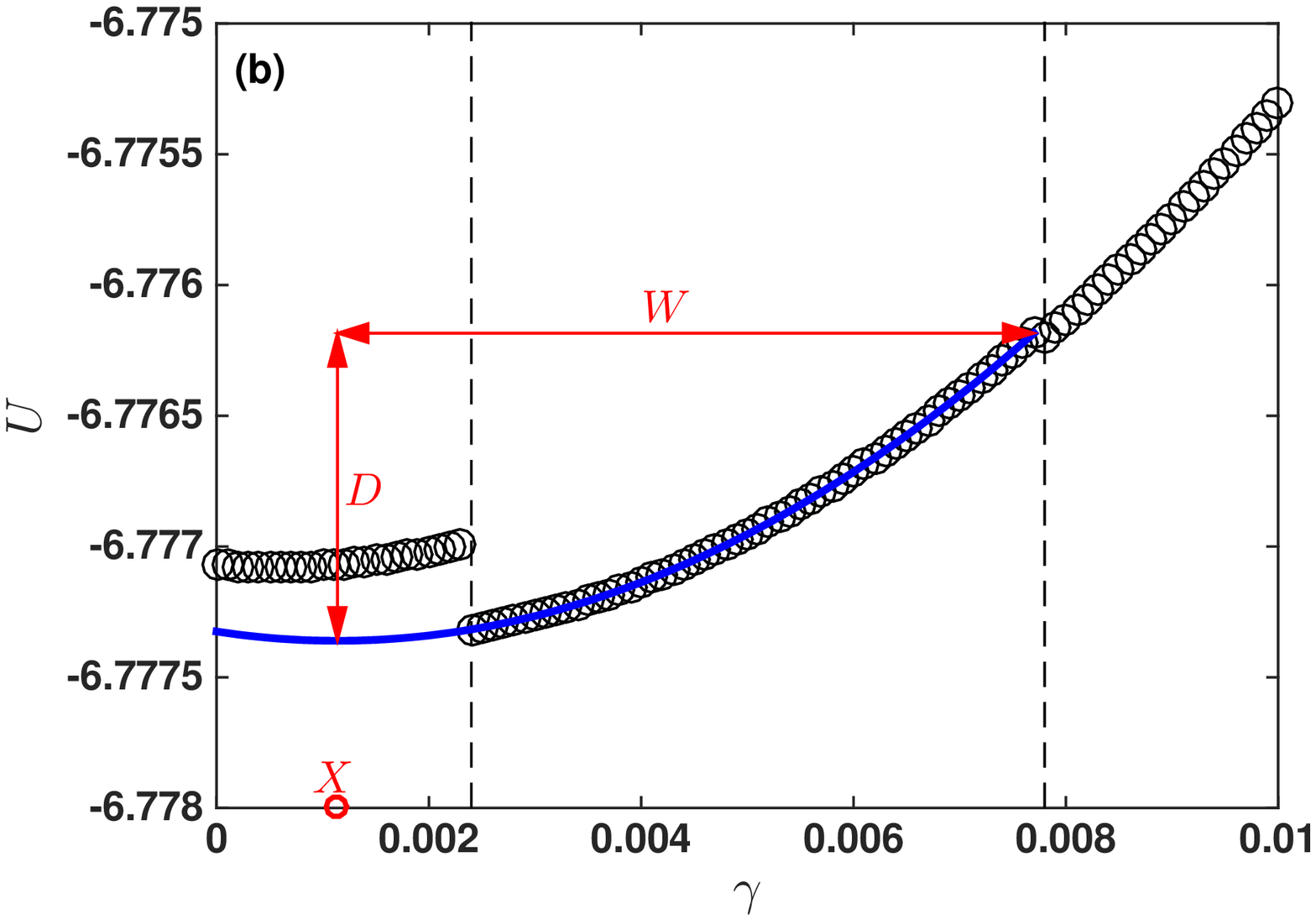}
\caption{(a) Potential energy $U$ versus strain $\gamma$ for a single glass
configuration (thick black curve) with $N=2000$ and prepared with 
cooling rate $R=10^{-2}$. 
The best fit parabolas for each of the continuous 
segments of $U(\gamma)$ in the range $0 < \gamma < 0.09$ are shown 
as thin blue curves. (b) Close-up of $U(\gamma)$ in the range $0 < \gamma < 0.01$ indicated by the red rectangle in panel (a).  
For this configuration (black circles) and range of strain, two 
rearrangements occur (indicated by dashed vertical lines). 
We show the best-fit parabola (solid blue curve) 
for the continuous segment between the two rearrangements. 
The half-width $W$ and depth $D$ of
the basin are indicated by the red arrows. The strain location of the 
minimum $\gamma=X$ of the continuous parabolic segment is given by the open 
circle.}
\label{fig:methodbasin}
\end{center}
\end{figure}

As shown in Fig.~\ref{fig:methodbasin} (a), the potential energy
versus strain $U(\gamma)$ for a single glass configuration is composed
of a series of continuous parabolic segments punctuated by rapid
rearrangement-induced drops. Along the continuous segments in strain,
the system remains in a series of similar minima in the potential
energy landscape. As the strain continues to increase, the potential
energy minimum will become unstable, the system will undergo a
rearrangement and move to a new minimum. With subsequent increases in
strain, the system will follow a new continuous parabolic segment
until that energy minimum becomes unstable. In
Fig.~\ref{fig:methodbasin} (b), we define several geometric features of
the PEL along the strain direction.  For each
continuous segment of $U(\gamma)$, we find the best fit parabola using
Eq.~\ref{parafit} with half-width $W=B/A + \gamma_i$, depth
$D=U(\gamma_i)-C +B^2/(2A)$, and strain location of the minimum
$X=-B/A$, where $\gamma_i$ is the strain at which a rearrangement
occurs (on the large strain side of the continuous segment).

\begin{figure}
\begin{center}
\includegraphics[width=1.0\columnwidth]{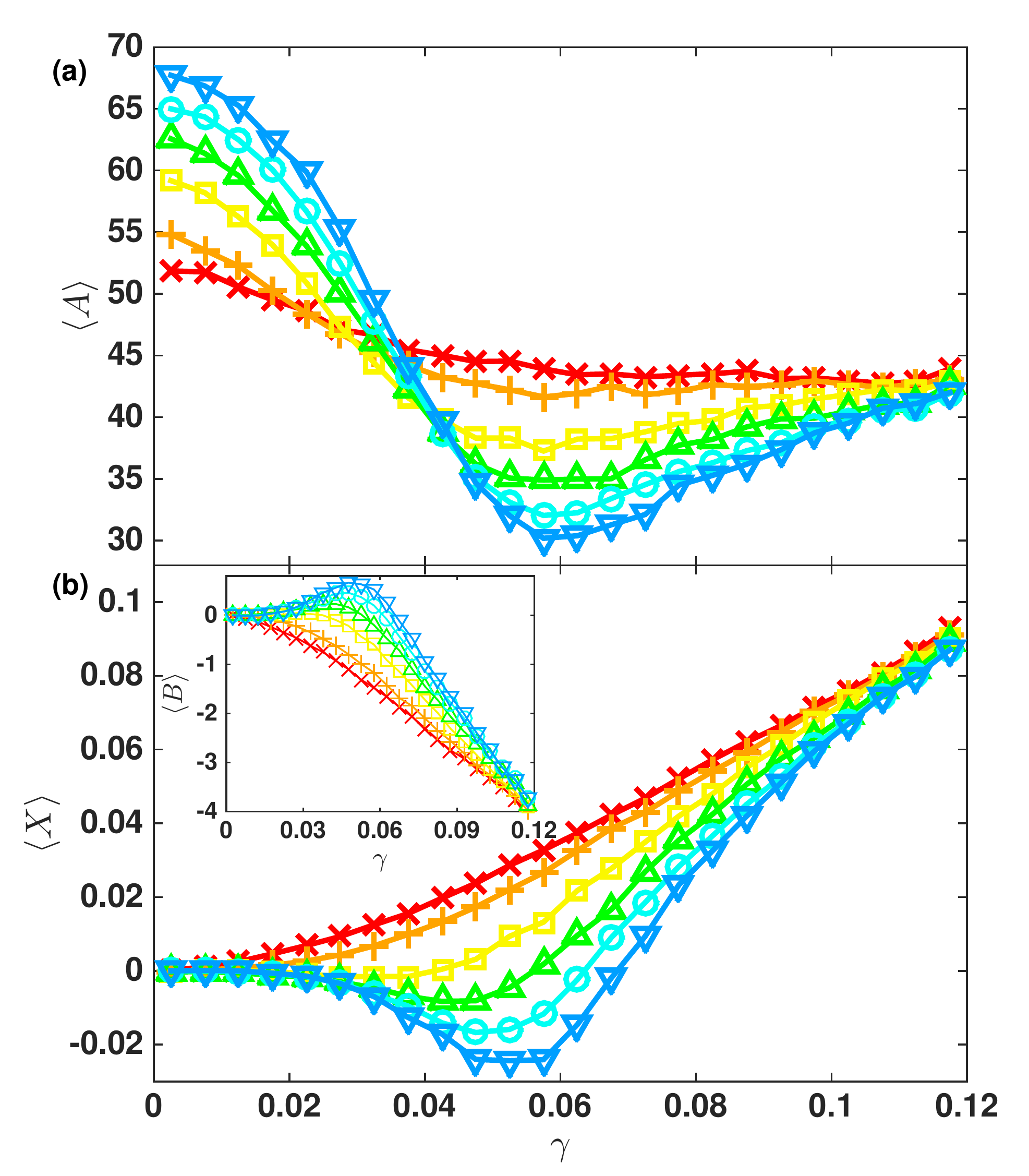}
\caption{Ensemble-averaged features of the PEL 
along the strain direction. We plot (a) the concavity $\langle A\rangle$ and 
(b) strain location of the potential energy minimum $\langle X\rangle$ for 
the continuous segments versus $\gamma$. In the inset, we also include $\langle B\rangle$ 
versus $\gamma$ for the continuous segments. For all data, we 
show six cooling rates, $R=10^{-1}$ (crosses), $10^{-2}$ (plus signs), 
$10^{-3}$ (squares), $10^{-4}$ (upward triangles), $10^{-5}$ (circles), 
and $10^{-6}$ (downward triangles), and average  
over $500$ samples with $N=2000$.}
\label{fig:subbasin1}
\end{center}
\end{figure}

In Fig.~\ref{fig:subbasin1}, we show the ensemble-averaged potential
energy landscape parameters $\langle A\rangle$, $\langle X\rangle$, and $\langle B\rangle$ as a function of strain
$\gamma$ and cooling rate $R$. Similar to the ensemble-averaged shear
modulus $\langle G\rangle$ in Fig.~\ref{fig:SS}, the concavity $\langle A\rangle$
depends weakly on $\gamma$ for rapidly cooled glasses. However, $\langle A\rangle$
becomes increasingly non-monotonic in $\gamma$ as the cooling rate
decreases. In panel (b), we show that the strain location of the basin
minimum occurs at $\langle X\rangle=0$ at $\gamma=0$, and $\langle X\rangle$ either increases with
$\gamma$ (for large $R$) or decreases with $\gamma$ (for small $R$)
depending on the cooling rate. Large deviations from $\langle X\rangle=0$ are
associated with yielding. For rapidly cooled glasses, there are many
nearby minima in the PEL~\cite{buchner1999potential} with similar values of
$\langle A\rangle$, $\langle B\rangle<0$, and values of $|\langle B\rangle|$ that increase with strain.  Thus,
rapidly cooled glasses possess basins with $\langle X\rangle$ that increase with
$\gamma$.  For more slowly cooled glasses, rearrangements below
yielding are less intense (Fig.~\ref{fig:RS} (c)). In this case, $\langle B\rangle$
changes signs and $\langle A\rangle$ decreases with strain near yielding.  As a
result, $\langle X\rangle < 0$ for slowly cooled glasses in the strain regime near
yielding.  At large $\gamma$, $\langle X\rangle \sim \gamma$ for all cooling rates.

There are two contributions to the potential energy loss $\langle U_{\rm
  loss}'\rangle$ from softening.  The first contribution, from the
integration of $(A_0-A(\gamma))\gamma$ over strain, is similar to the
stress loss from softening $\langle \sigma_{\rm loss}'\rangle$. The second
contribution stems from the integration of $B_0-B(\gamma)$ over
$\gamma$.  For rapidly cooled glasses, the second contribution to
$\langle U_{\rm loss}'\rangle$ is larger than the first for all strains.  For slowly
cooled glasses, when $\langle B(\gamma)\rangle$ becomes sufficiently positive near
yielding (inset to Fig.~\ref{fig:subbasin1} (b)), the second
contribution can switch from positive to negative, providing an
effective potential energy gain.  However, for slowly cooled glasses,
the potential energy loss from the first contribution is much larger
than the effective gain, and thus $\langle U_{\rm loss}'(\gamma)\rangle$ also grows
with $\gamma$ for slowly cooled glasses as shown in
Fig.~\ref{fig:energy} (a).

\begin{figure}[!ht]
\begin{center}
\includegraphics[width=0.9\columnwidth]{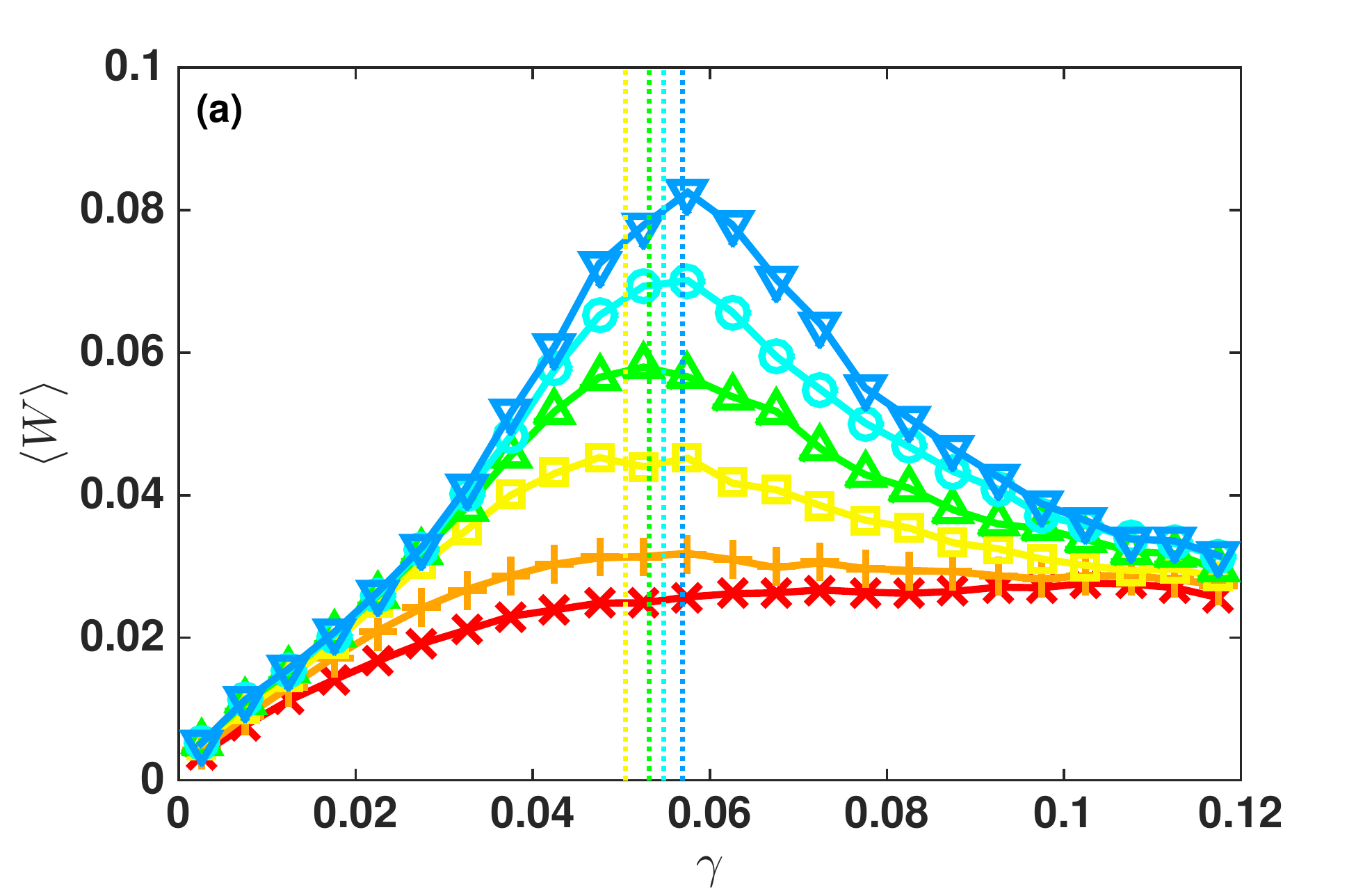}
\includegraphics[width=0.9\columnwidth]{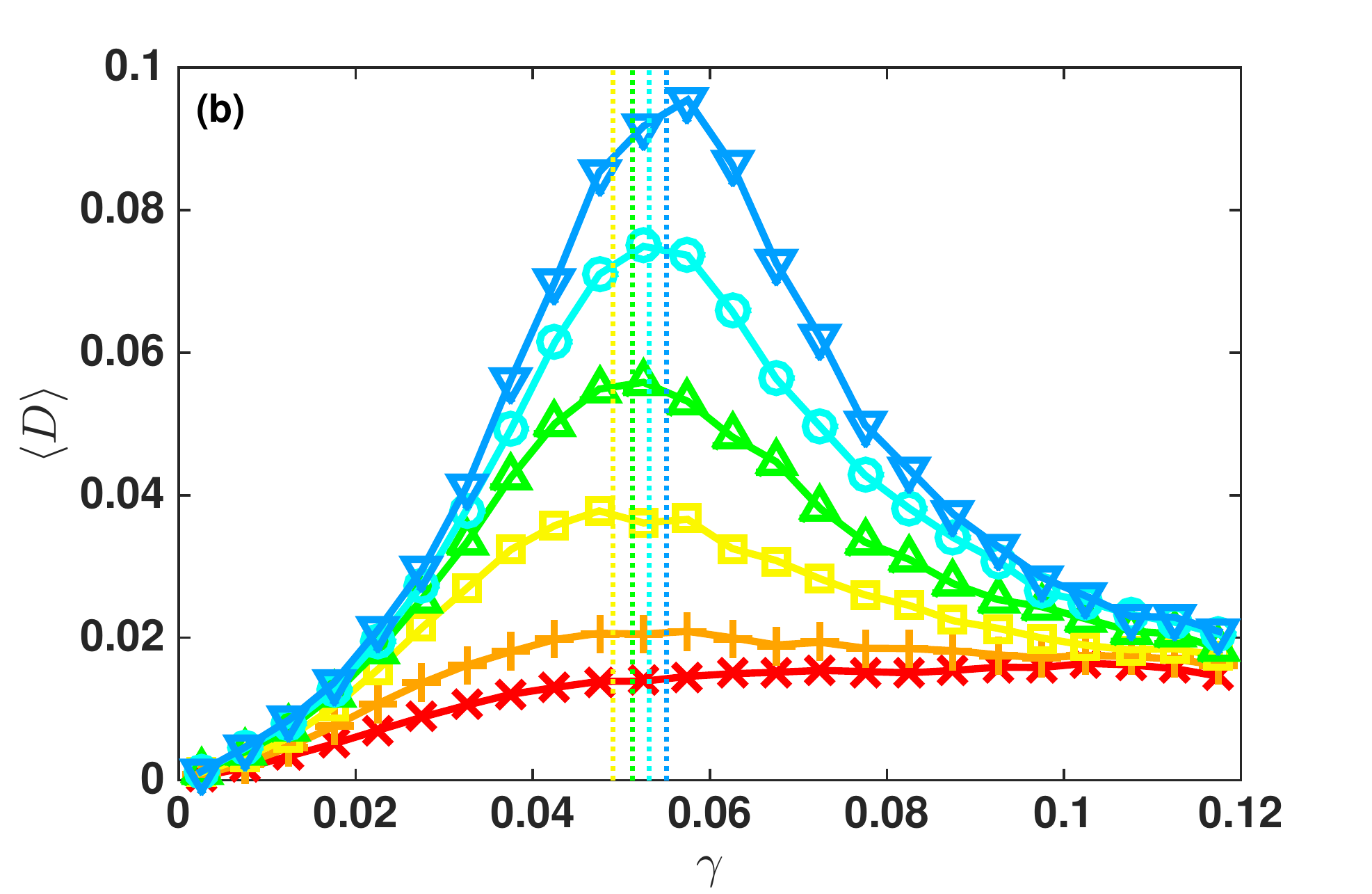}
\includegraphics[width=0.9\columnwidth]{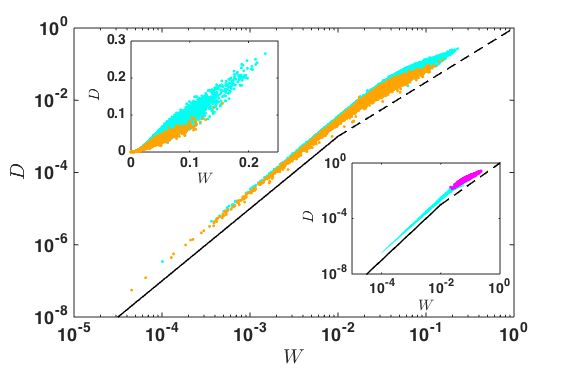}
\caption{Ensemble-average of (a) the half-width $\langle W\rangle$ and (b) depth $\langle D\rangle$
of the continuous segments of the potential energy $U(\gamma)$
versus the midpoint strain for each segment for cooling rates $R=10^{-1}$ (crosses), $10^{-2}$ (plus signs), $10^{-3}$
(squares), $10^{-4}$ (upward triangles), $10^{-5}$ (circles), and $10^{-6}$ (downward triangles).
The strains $\gamma^*$ at which $\langle W\rangle$ and $\langle D\rangle$ form a peak are indicated 
by dotted lines for those cooling rates $R$ where a peak is clearly visible.
(c) Scatter plot of $D$ versus $W$ for all continuous 
segments in the strain interval $0 < \gamma < 0.12$ for $R=10^{-2}$ (orange) and $10^{-5}$ (cyan). The solid and dashed lines have slopes $2$ and $1.5$, 
respectively. The upper-left inset shows $D$ versus $W$ on a linear-linear 
scale for $R=10^{-2}$ (orange) and $10^{-5}$ (cyan), and the lower-right inset shows $D$ versus $W$ on a $\log_{10}$-
$\log_{10}$ scale for $R=10^{-5}$. Data near yielding 
($0.045 < \gamma < 0.065$) are colored magenta.}
\label{fig:subbasin2}
\end{center}
\end{figure}

We now focus on the strain and cooling rate dependence of the
half-width $W$ and depth $D$ of the basins that are sampled in the
PEL along the strain direction during AQS pure shear. As shown in Fig.~\ref{fig:subbasin2} (a) and
(b), the ensemble-averaged $\langle W\rangle$ and $\langle D\rangle$
possess similar dependence on strain and cooling rate.  $\langle
W\rangle=\langle D\rangle=0$ at $\gamma=0$ and then both increase with
$\gamma$ for small strains.  As $\gamma$ continues to increase,
$\langle W\rangle$ and $\langle D\rangle$ become cooling-rate
dependent. For rapidly cooled glasses, $\langle W\rangle$ and $\langle D\rangle$ grow monotonically
with strain, reaching plateau values ($\langle W\rangle \sim 0.03$ and $\langle D\rangle \sim 0.02$)
in the large-strain limit. In contrast, for slowly cooled glasses, $\langle W\rangle$
and $\langle D\rangle$ form peaks near $\gamma^* \sim 0.055$ before reaching their
large-strain plateau values. The values of $\gamma^*$ for slowly
cooled glasses are similar to those for the peak locations of the von
Mises stress $\langle \sigma(\gamma)\rangle$ (Fig.~\ref{fig:stress} (b)), which
indicates that as the strain increases above yielding, the basin
geometries change dramatically. 

In Fig.~\ref{fig:subbasin2} (c), we show a scatter plot of $D$ versus
$W$ for all of the continuous parabolic segments in $U(\gamma)$ in the
range $0 < \gamma < 0.12$.  We find that more slowly cooled glasses
sample basins with larger depths and half-widths, $D$ and $W$, as
shown in the upper left inset to panel (c).  At small strains, and for all
cooling rates, the half-width of the basins scales quadratically with
the depth, $W\sim D^2$~\cite{fan2014thermally}. In contrast, $W\sim
D^{\lambda}$ with $\lambda \sim 1.5$ at large strains near and above
yielding, which signifies that the dynamics has transitioned from
intra-metabasin to inter-metabasin
sampling~\cite{johnson2005universal,maloney2006energy}. (See the lower right inset to
panel (c).)  Recent studies of unsheared, finite-temperature glasses
have also shown that the basin widths and depths are larger for more
slowly cooled glasses. However, these studies also showed that the
basin curvature is independent of cooling rate, which differs from the
results presented in Fig.~\ref{fig:subbasin1} (a) for glasses
undergoing AQS pure shear.  Thus, thermal
fluctuating systems and glasses undergoing AQS pure
shear sample basins with different geometric properties.  In summary,
we have shown that the geometric properties of basins in the potential
energy landscape vary strongly near yielding and depend strongly on
cooling rate for glasses undergoing AQS pure shear.

\subsection{Yielding transition}
\label{sec:results_Neffect}

\begin{figure}
\begin{center}
\includegraphics[width=0.9\columnwidth]{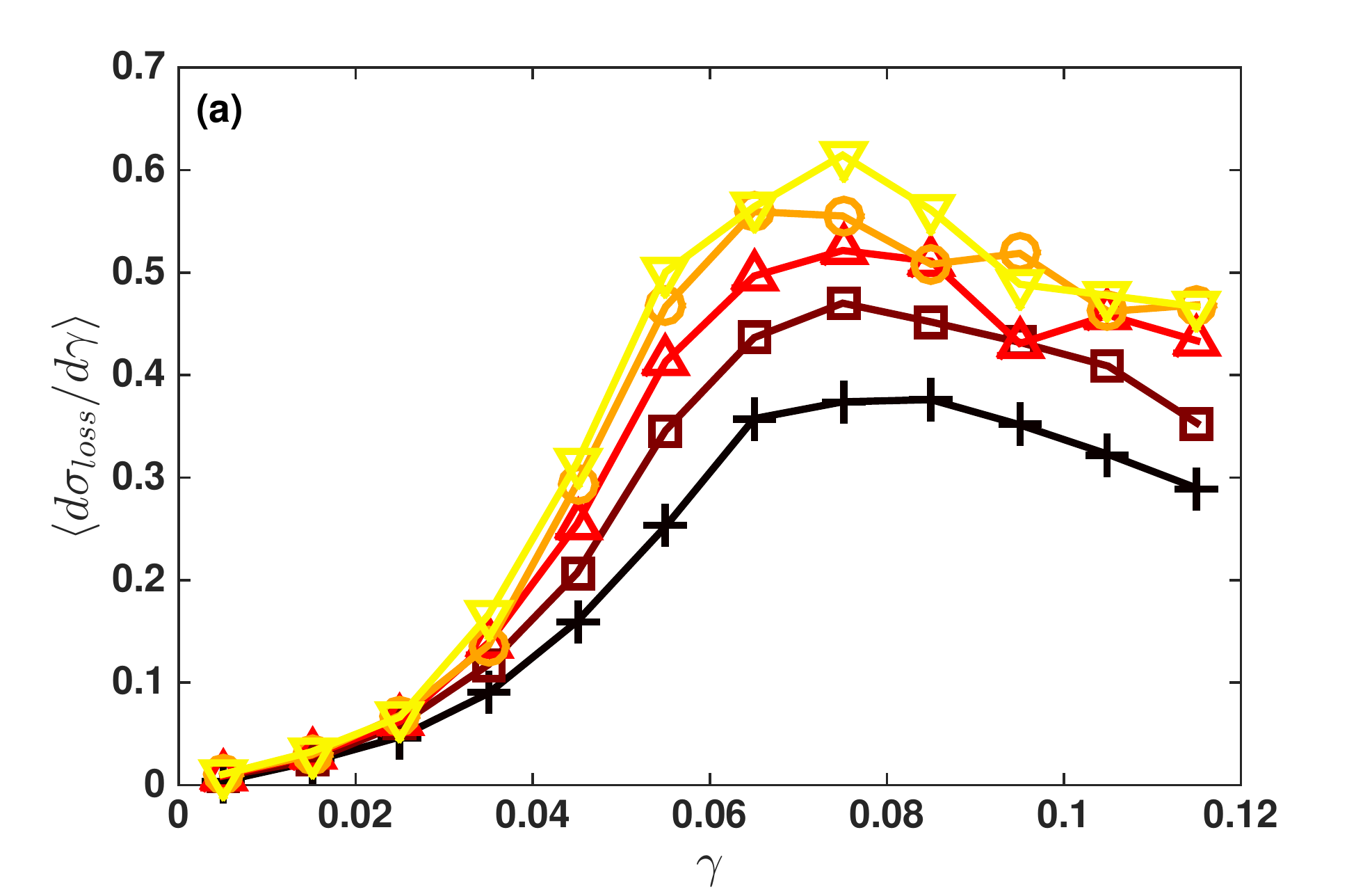}
\includegraphics[width=0.9\columnwidth]{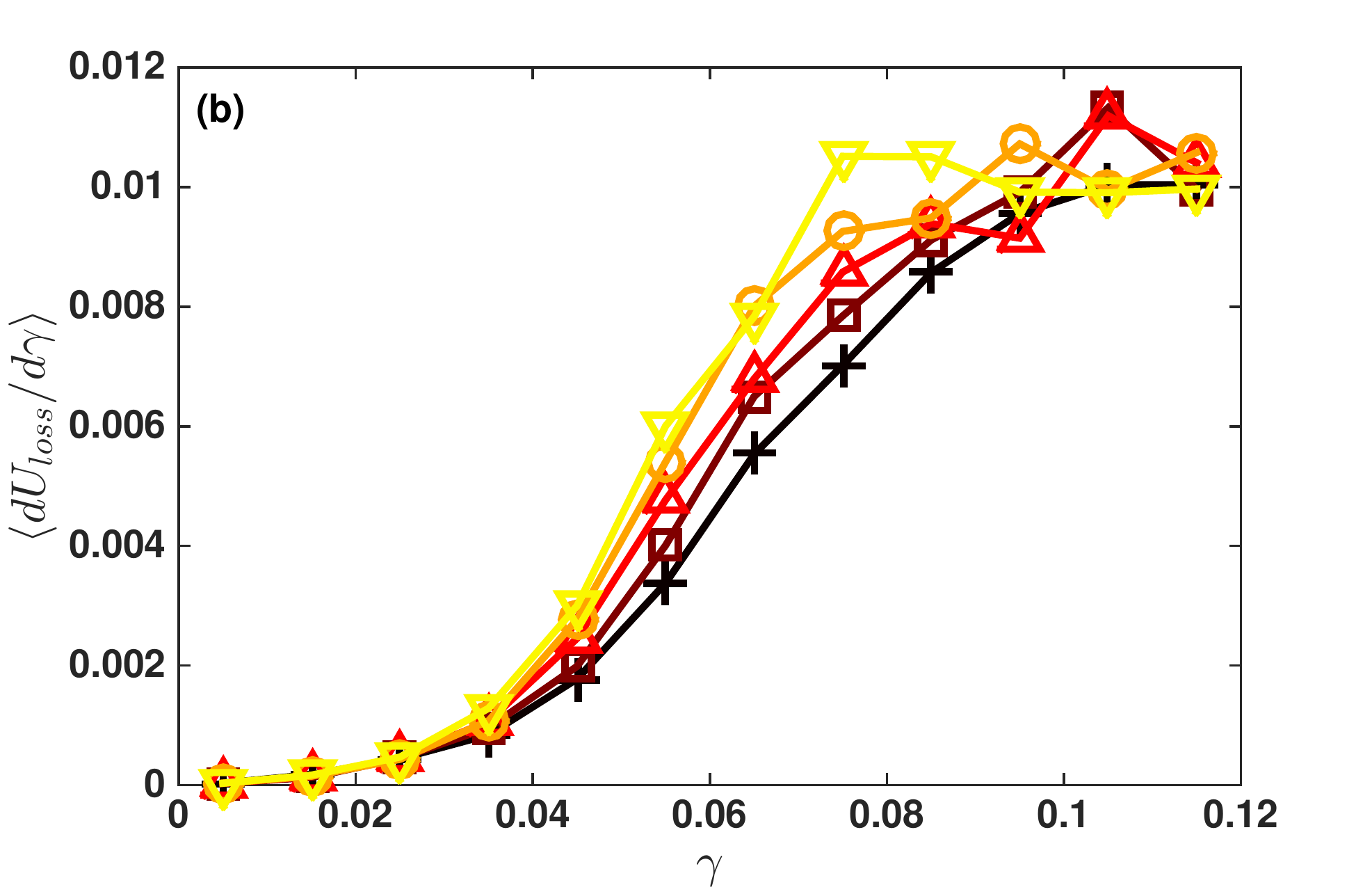}
\caption{The ensemble-averaged (a) rearrangement-induced stress loss per (1\%) strain
$\langle d\sigma_{\rm loss}/d\gamma\rangle$ and (b) rearrangement-induced energy loss per (1\%) strain
$\langle dU_{\rm loss}/d\gamma \rangle$ plotted versus strain $\gamma$ for glasses prepared
at cooling rate $R=10^{-5}$ and several system sizes: $N=250$ (crosses), $500$ (squares),
$1000$ (upward triangles), $2000$ (circles), and $4000$ (downward triangles).
All data points are obtained by averaging over at least $500$ independent samples.
}
\label{fig:N_RS}
\end{center}
\end{figure}

In this section, we analyze the system-size dependence of the stress
and energy losses from rearrangements and softening.  Prior studies
have shown that quantities, such as the average energy drop and
participation number during rearrangements, scale sublinearly with
system size in glasses undergoing AQS
shear~\cite{maloney2004subextensive,karmakar2010statistical}.  Other
work has shown that changes in the scaling of the rearrangement
statistics with system size are associated with the yielding
transition~\cite{hentschel2015stochastic,leishangthem2017yielding}.

\begin{figure}
\begin{center}
\includegraphics[width=0.9\columnwidth]{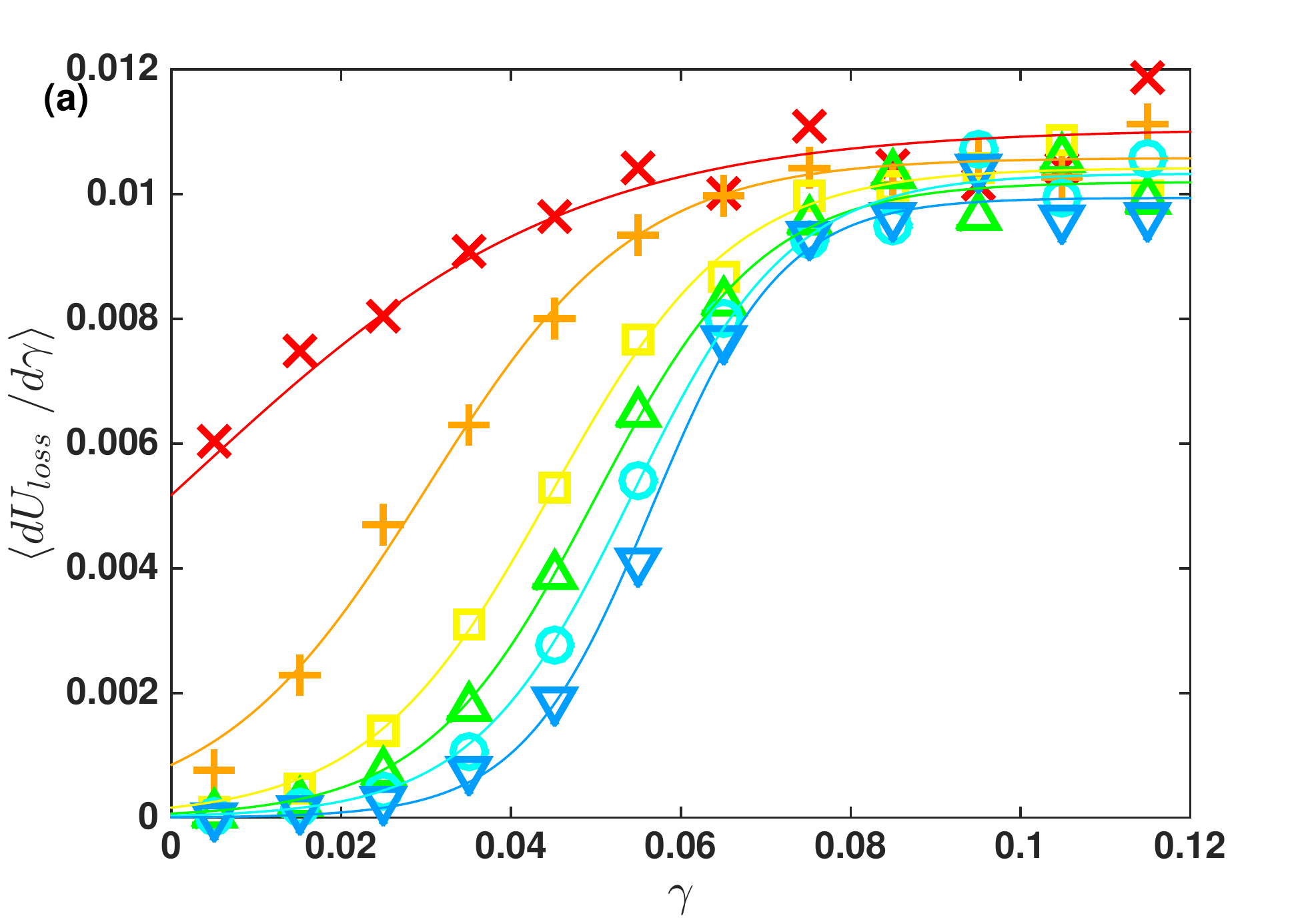}
\includegraphics[width=0.9\columnwidth]{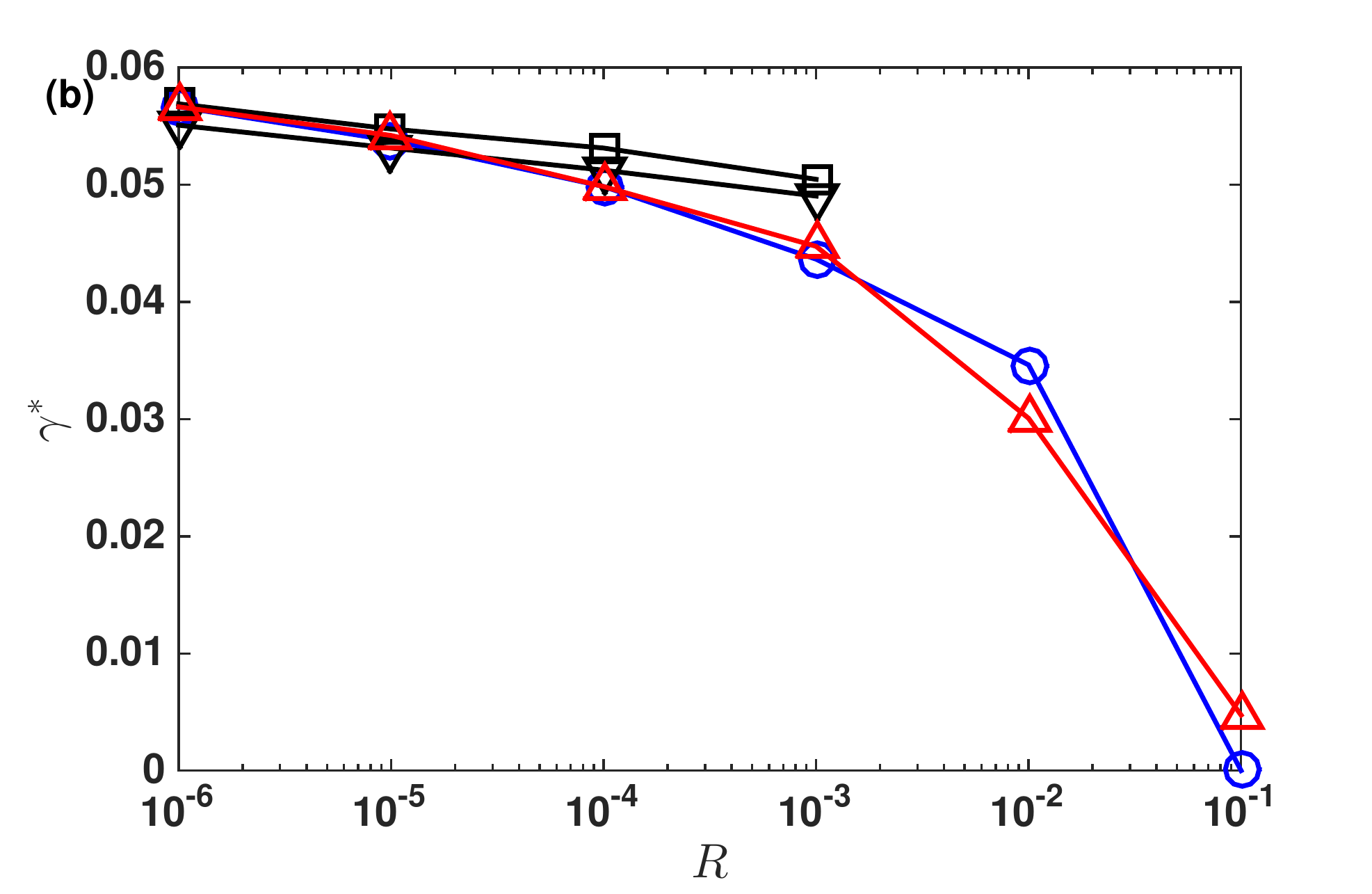}
\caption{(a) The ensemble-averaged
rearrangement-induced energy loss per (1\%) strain $\langle dU_{\rm
loss}/d\gamma\rangle$ plotted versus strain $\gamma$ for glasses with $N=2000$
and prepared at several cooling rates: $R=10^{-1}$ (crosses), $10^{-2}$ (plus
signs), $10^{-3}$ (squares), $10^{-4}$ (upward triangles), $10^{-5}$
(circles), and $10^{-6}$ (downward triangles).  $\langle dU_{\rm loss}/d\gamma \rangle$ 
is obtained by averaging over $500$ independent samples.  The solid 
curves are the best-fit logistic functions for each $R$.
(b) Several characteristic strains $\gamma^*$ plotted versus cooling rate $R$ for glasses with $N=2000$. We plot the inflection points for the potential 
energy $\langle U(\gamma)\rangle$ (circles) multiplied by $\approx 1.3$ and potential energy loss per strain $\langle dU_{\rm loss}/d\gamma\rangle$ (upward triangles) and the location of the peaks in the 
half-width $\langle W\rangle$ (squares) and 
depth $\langle D\rangle$ (downward triangles) of basins in the potential energy 
landscape for small $R$.}
\label{fig:ys}
\end{center}
\end{figure}

First, note that macroscale quantities, such as the ensemble-averaged
stress $\sigma(\gamma)$ and potential energy per particle $U(\gamma)$,
are largely independent of system size for $N \gtrsim 500$.
(See Fig.~\ref{fig:app_N1} in Appendix~\ref{app:N}.)  In
Fig~\ref{fig:N_RS}, we show the system-size dependence of the
rearrangement-induced stress loss per strain $\langle d\sigma_{\rm
  loss}(\gamma)/d\gamma \rangle$ and energy loss per strain $\langle dU_{\rm
  loss}(\gamma)/d\gamma \rangle$. For $N \gtrsim  1000$, $\langle d\sigma_{\rm
  loss}(\gamma)/d\gamma \rangle$ and $\langle dU_{\rm loss}(\gamma)/d\gamma \rangle$ are
nearly independent of system size at small and large strains.
However, at strains near the yield strain $\gamma_y\sim0.055$, both
$\langle d\sigma_{\rm loss}(\gamma)/d\gamma \rangle$ and $\langle dU_{\rm
  loss}(\gamma)/d\gamma \rangle$ display sharper increases with strain as $N$
increases.  For slowly cooled glasses, $\langle d\sigma_{\rm
  loss}(\gamma)/d\gamma \rangle$ forms a peak near yielding ({\it cf.}
Fig.~\ref{fig:RS} (c)), which persists in the large system limit.  In
contrast, $\langle dU_{\rm loss}(\gamma)/d\gamma \rangle$ does not possess a peak and
instead displays a sigmoidal form for all cooling
rates~\cite{fan2017effects}. The slope of $\langle dU_{\rm
  loss}(\gamma)/d\gamma \rangle$ near the midpoint of the sigmoid sharpens
with increasing $N$, but reaches a (cooling-rate dependent) finite
value in the large-system limit.  The large-system limit for the slope
of $\langle dU_{\rm loss}/d\gamma \rangle$ (near the midpoint) grows with decreasing
cooling rate. (See Fig.~\ref{fig:ys} (a).)
The rapid increase in the slope of $\langle d U_{\rm loss}/d\gamma \rangle$ signals a
significant acceleration of rearrangements and energy loss near the
yielding transition.

\begin{figure}
\begin{center}
\includegraphics[width=0.9\columnwidth]{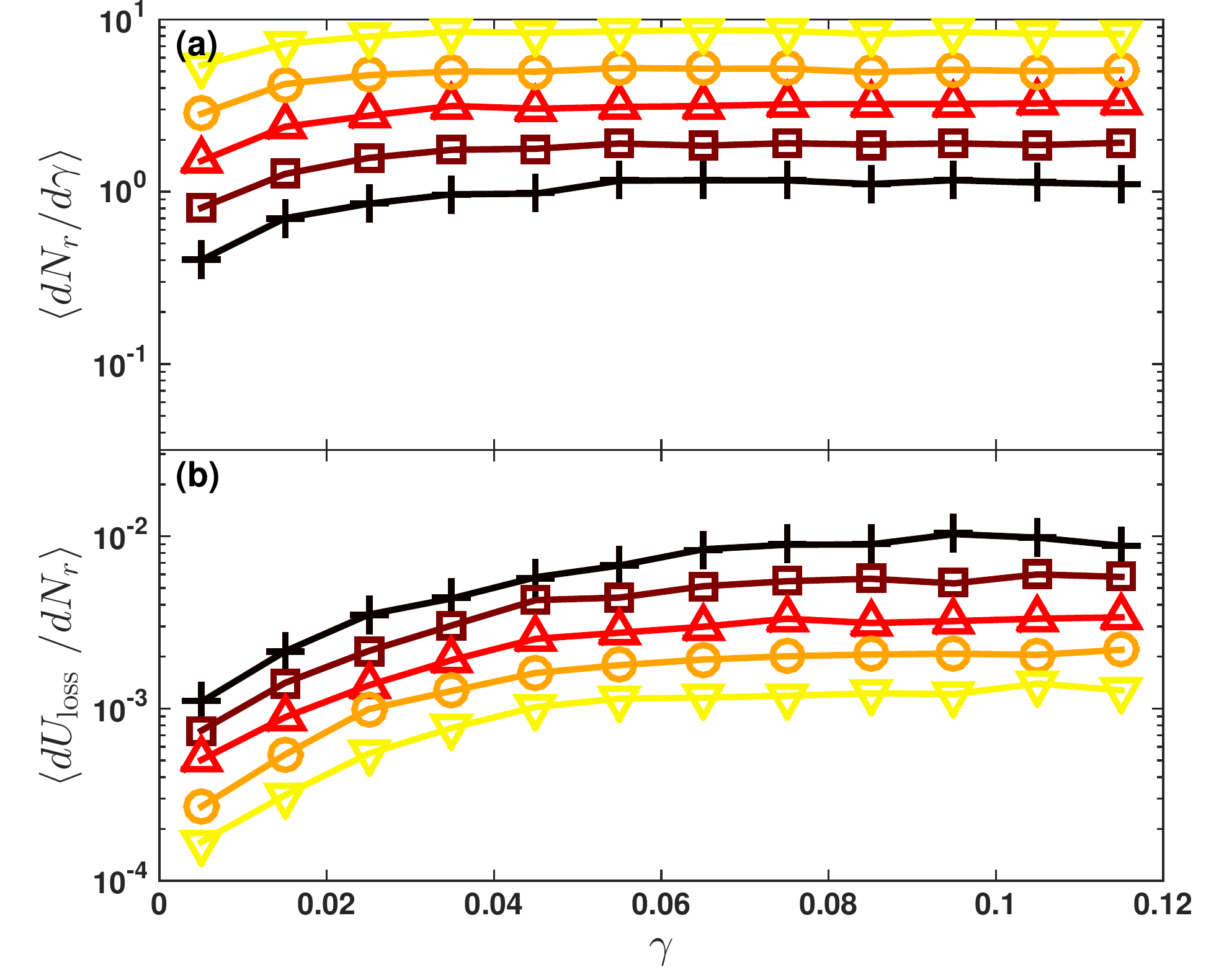}
\caption{The ensemble-averaged (a) rearrangement frequency $\langle dN_r/d\gamma \rangle$ 
and (b) energy loss per rearrangement $\langle dU_{\rm loss}/dN_r \rangle$ plotted versus
strain $\gamma$ for glasses prepared with cooling rate $R=10^{-2}$
and several system sizes: $N=250$ (crosses), $500$ (squares), $1000$ 
(upward triangles), $2000$ (circles), and $4000$ (downward triangles).
All data points are obtained by averaging over at least $500$ samples.}
\label{fig:freqsize}
\end{center}
\end{figure}

\begin{figure}
\begin{center}
\includegraphics[width=0.9\columnwidth]{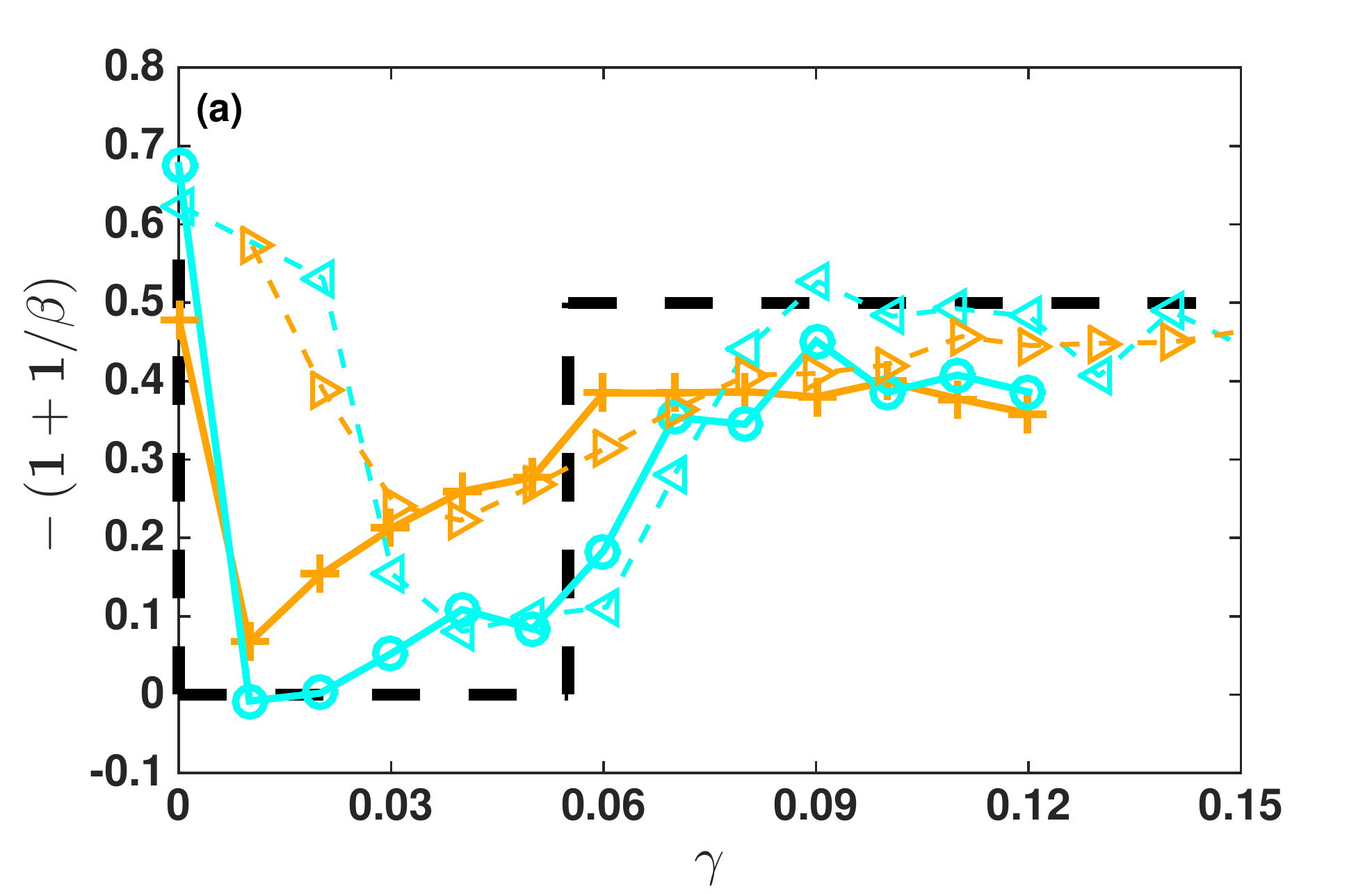}
\includegraphics[width=0.9\columnwidth]{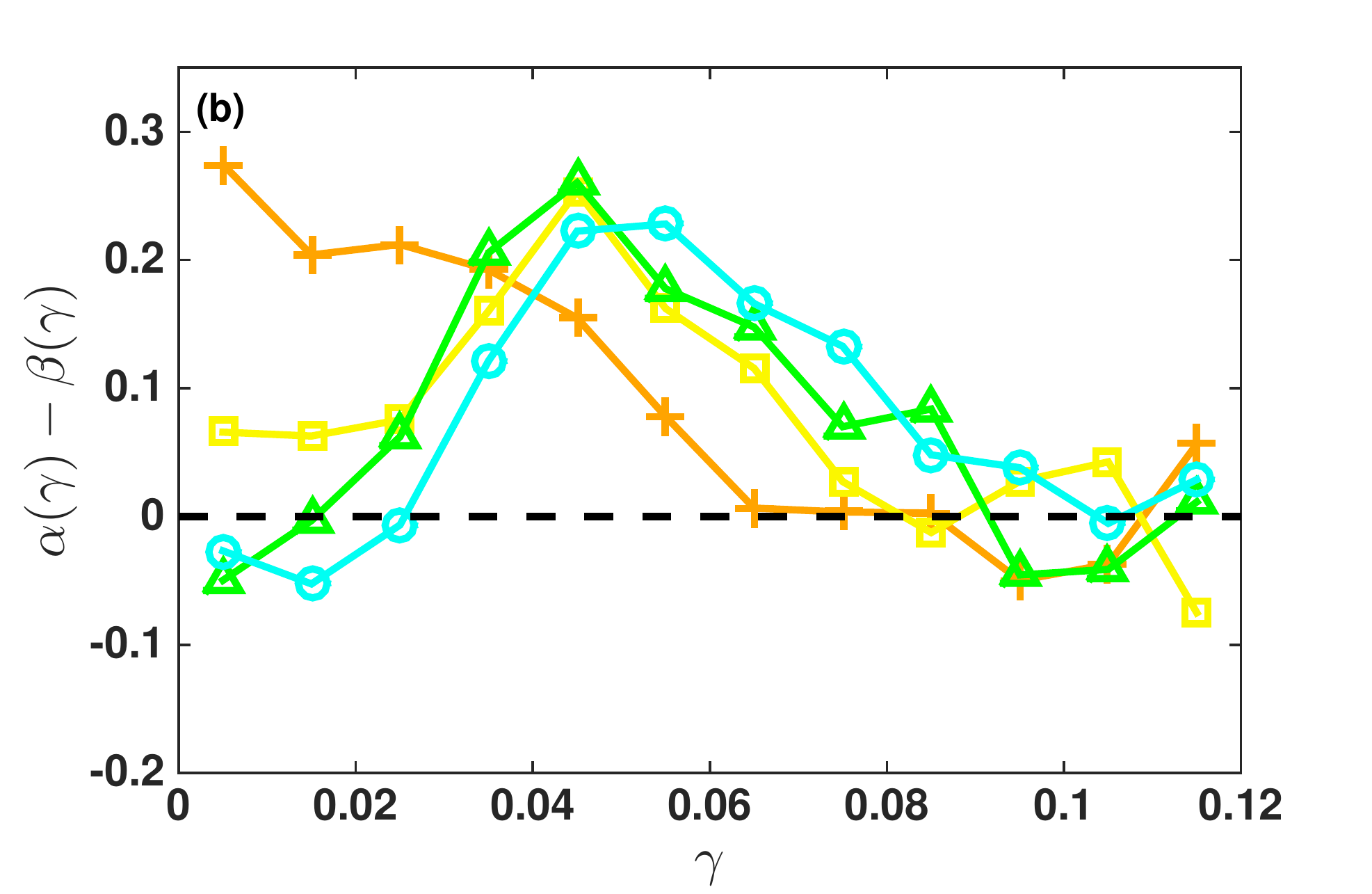}
\caption{(a) The quantity $-(1+1/\beta(\gamma))$ plotted as a function
of strain $\gamma$ and two cooling
rates, $R=10^{-2}$ (crosses) and $10^{-5}$ (circles), where the rearrangement frequency $\langle dN_r/d\gamma\rangle$
scales as a power law with system size with exponent $\beta$: $\langle dN_r/d\gamma \rangle \sim
N^{-\beta}$.  We also show data for $-(1+1/\beta)$ from Hentschel, {\it et
al.}~\cite{hentschel2015stochastic} (dashed curves) with 
``infinitely fast'' (rightward triangles) and slow
$R=10^{-5}$ (leftward triangles) cooling rates.
The theoretical 
prediction for $(-1+1/\beta)$ from Hentschel, {\it et
al.}~\cite{hentschel2015stochastic}, which is indicated by the black dashed
line, has an abrupt increase at the yielding transition.  (b) The difference between the scaling 
exponents $\alpha(\gamma)-\beta(\gamma)$ is plotted as a function of strain
$\gamma$ for cooling rates $R=10^{-2}$
(crosses), $10^{-3}$ (squares), $10^{-4}$ (triangles), and $10^{-5}$
(circles). $\alpha(\gamma)$ is the system-size scaling exponent for the energy loss 
per rearrangement: $\langle dU_{\rm loss}/dN_r\rangle \sim N^{\alpha(\gamma)}$.}
\label{fig:scaling}
\end{center}
\end{figure}

As described in Sec.~\ref{sec:results_stress} for $\sigma_{\rm loss}$,
we can decompose $dU_{\rm loss}(\gamma)/d\gamma$ into two
contributions that give the size $dU_{\rm loss}(\gamma)/dN_r$ and
frequency $dN_r/d\gamma$ of rearrangements.  In
Fig.~\ref{fig:freqsize}, we show the system size dependence of the ensemble 
average of these
two quantities.  As $N$ increases, the rearrangement size decreases
and the frequency increases.  Previous
studies~\cite{hentschel2015stochastic,lerner2009locality,karmakar2010statistical}
have focused on the system-size scaling of similar quantities: 1) the
strain interval $\Delta\gamma \sim (dN_r/d\gamma)^{-1}$ between
rearrangements and 2) the total energy loss per rearrangement
$\Delta{\cal U} \sim NdU_{\rm loss}/dN_r$.

The ensemble-averaged size and frequency of rearrangements display
power-law scaling with system size:
\begin{equation}
\label{eq:scaling1}
\langle dU_{\rm loss}/dN_r\rangle \sim N^{\alpha(\gamma)}
\end{equation}
\begin{equation}
\label{eq:scaling2}
\langle dN_r/d\gamma \rangle \sim N^{-\beta(\gamma)},
\end{equation}
where the scaling exponents $\alpha(\gamma)$ and $\beta(\gamma)$ are
functions of strain $\gamma$ and cooling rate $R$.  In
Fig~\ref{fig:scaling} (a), we compare our results for
$-(1+1/\beta(\gamma))$ with those from
Ref.~\cite{hentschel2015stochastic} for several cooling rates.
Ref.~\cite{hentschel2015stochastic} provided theoretical arguments for
the strain dependence of $-(1+1/\beta(\gamma))$. They argued that
$-(1+1/\beta(\gamma))$ should jump from a nonzero, non-universal value
($\approx 0.6$ for binary Lennard-Jones glasses) at $\gamma=0$ to $0$
when $\gamma >0$, then jump discontinuously from zero to a nonzero
value at the yield strain $\gamma_y$, and remain at a universal value
$0.5$ as the strain increases beyond $\gamma_y$.  As shown in
Fig~\ref{fig:scaling} (a), our data is qualitatively similar to the
data for Ref.~\cite{hentschel2015stochastic}.  In particular,
$-(1+1/\beta(\gamma))$ decreases from a maximal value at $\gamma=0$,
remains roughly constant and small over a narrow strain interval below
the yield strain, and then begins to increase beyond the yield strain,
approaching a plateau value near $0.5$ at large strains.

The data in Fig~\ref{fig:scaling} (a) suggests that $-(1+1/\beta)$
decreases with decreasing cooling rate in the range $0.04 < \gamma <
0.07$, but it does not depend strongly on the cooling rate at large
strains. Much larger ensemble averages should be performed to confirm
these results.  Using Eqs.~\ref{eq:scaling1} and~\ref{eq:scaling2},
the rearrangement-induced energy loss per strain obeys $\langle dU_{\rm
  loss}/d\gamma\rangle \sim N^{\alpha(\gamma)-\beta(\gamma)}$.  In
Fig~\ref{fig:scaling} (b), we show that the difference in the scaling
exponents $\alpha(\gamma)-\beta(\gamma) \sim 0$ at small and large
strains, while $\alpha(\gamma)-\beta(\gamma)>0$ near the yield
strain. A positive value for $\alpha(\gamma) - \beta(\gamma)$
indicates that $\langle dU_{\rm loss}/d\gamma \rangle$ can serve as an order parameter
for the yielding transition.  The data for
$\alpha(\gamma)-\beta(\gamma)$ for rapidly cooled glasses with
$R=10^{-2}$ differs from that for more slowly cooled glasses.
The stress $\langle \sigma(\gamma)\rangle$ and stress loss from rearrangements
$\langle d\sigma_{\rm loss}/d\gamma\rangle$ do not possess peaks for rapidly cooled
glasses and, in this case, the yield transition behaves as a smooth 
crossover~\cite{hentschel2015stochastic}.

In Fig.~\ref{fig:ys} (b), we plot several characteristic strains
(inflection points in $\langle U(\gamma)\rangle$ (Fig.~\ref{fig:energy} (b)) and
$\langle dU_{\rm loss}/d\gamma \rangle$ (Fig.~\ref{fig:ys} (a)) and the peak locations
of the half-width $\langle W\rangle$ and depth $\langle D\rangle$ (Fig.~\ref{fig:subbasin2} (a) and
(b)) of the basins in the PEL), which are
correlated with the yielding transition, as a function of cooling
rate. At low cooling rates, these measures approach $\gamma^* \approx
0.055$. As the cooling rate increases, these characteristic strains
decrease.  In particular, the measures of the inflection points tend to zero
near $R_c \approx 10^{-1}$.  Note that $\langle W\rangle$ and $\langle D\rangle$ do not possess
peaks for cooling rates $R>10^{-3}$, and thus these data points are
not plotted.

\begin{figure}
\begin{center}
\includegraphics[width=0.9\columnwidth]{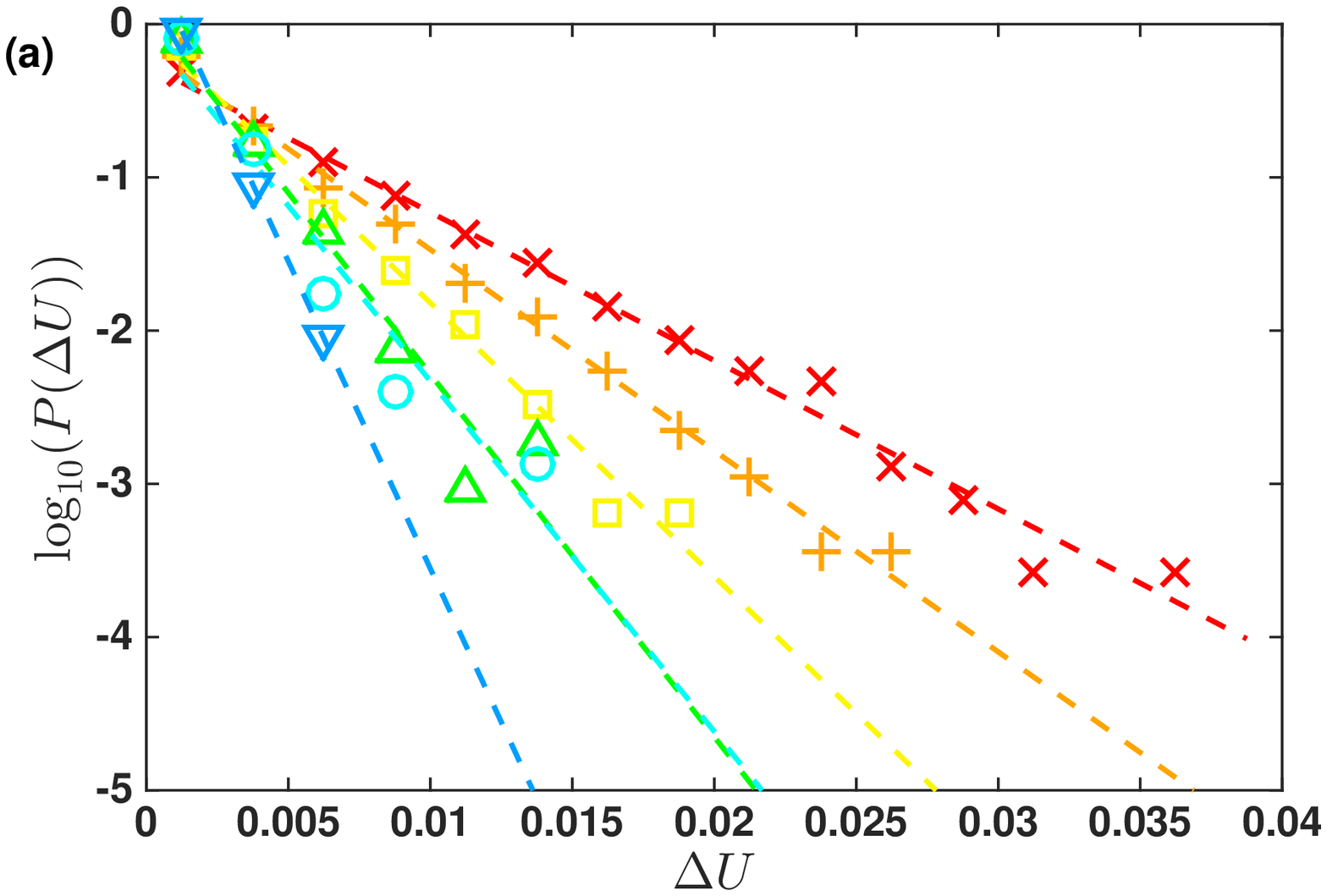}
\includegraphics[width=0.9\columnwidth]{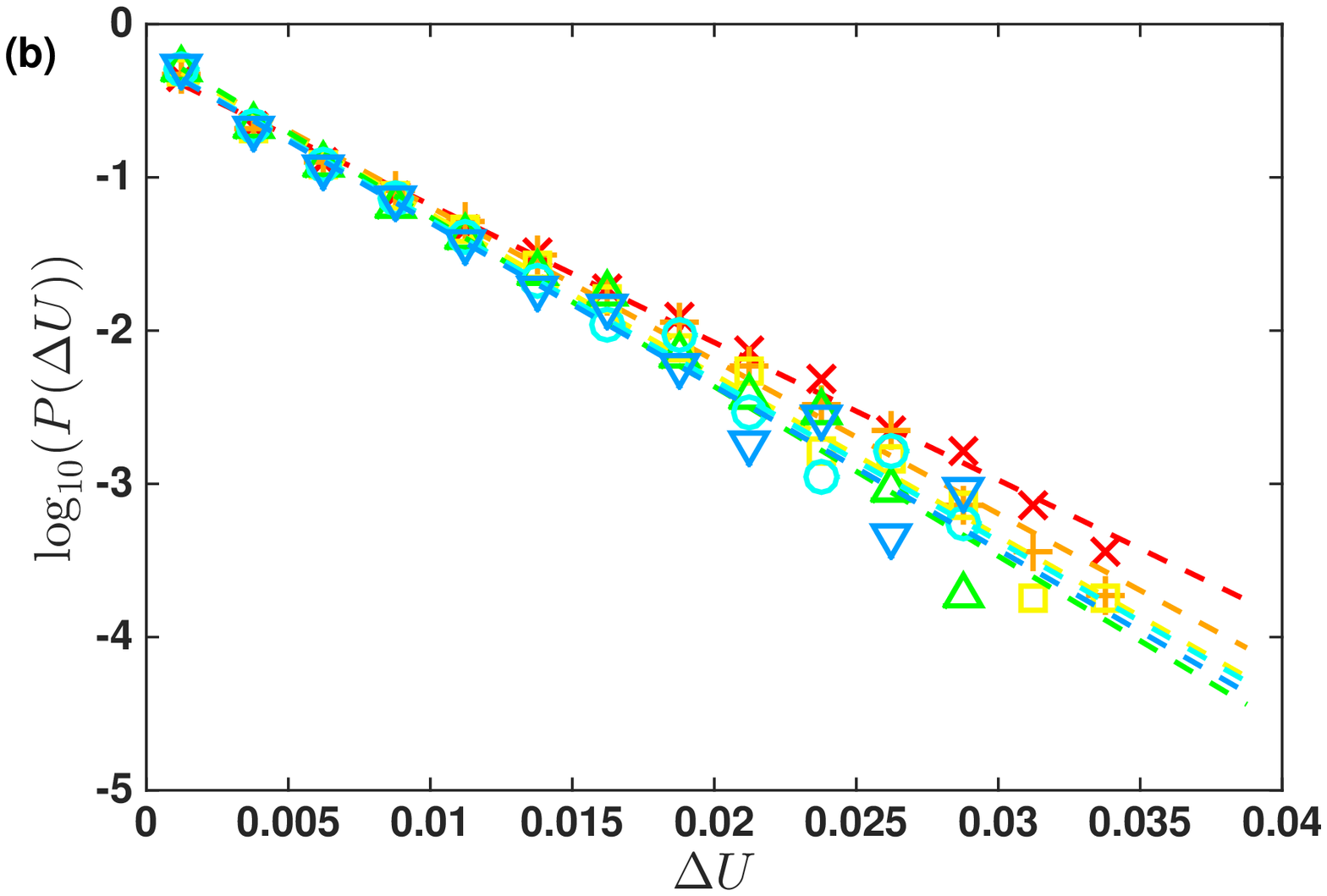}
\includegraphics[width=0.9\columnwidth]{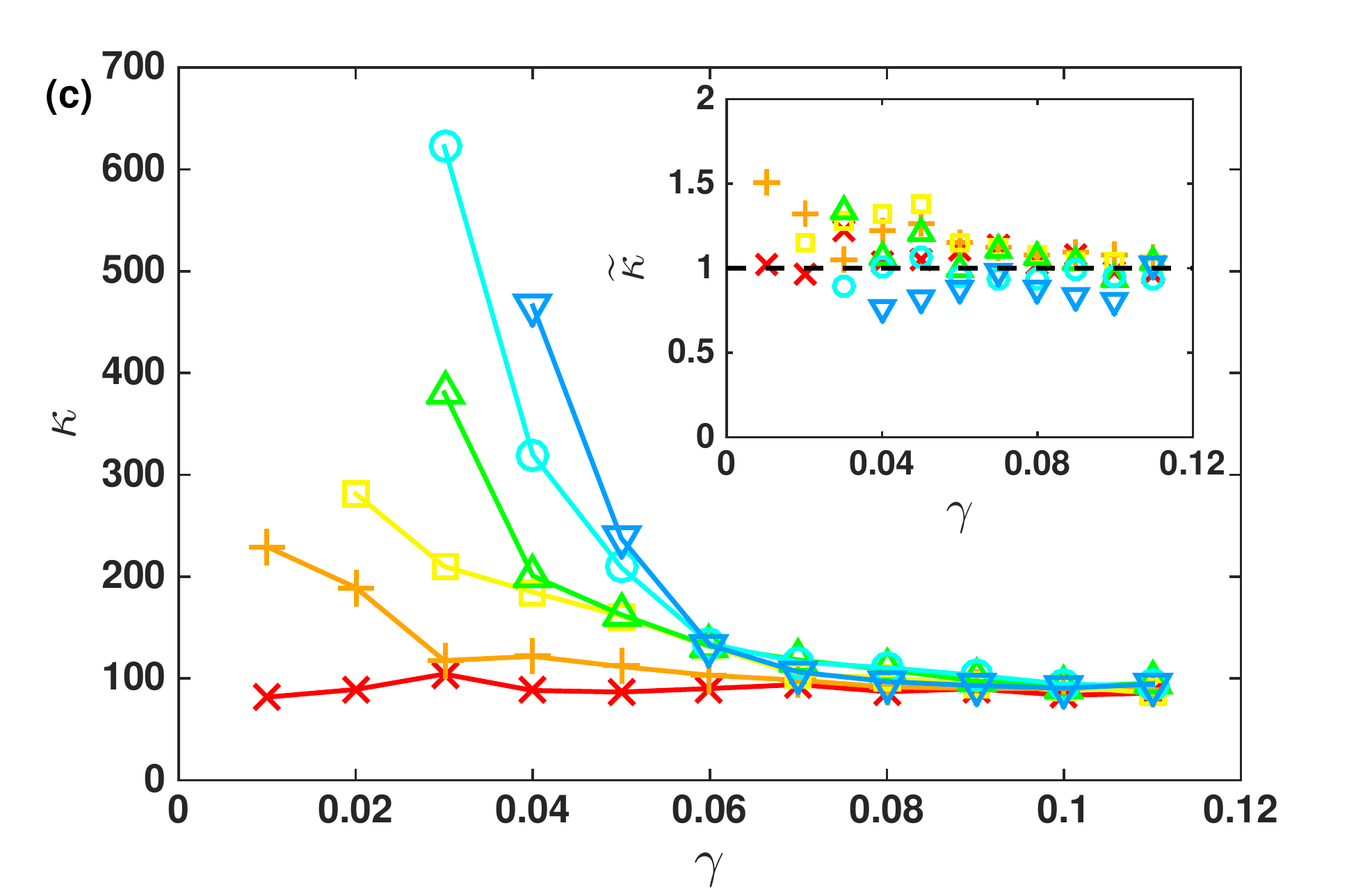}
\caption{The probability distribution $P(\Delta U)$ of energy drops 
(a) before ($\gamma<0.055$) and (b) after yielding 
($0.055<\gamma<0.12$) for glasses with $N=2000$ and prepared at 
cooling rates $R=10^{-1}$ (crosses), $10^{-2}$ (plus signs), $10^{-3}$
(squares), $10^{-4}$ (upward triangles), $10^{-5}$ (circles), and 
$10^{-6}$ (downward triangles). The distributions 
have been normalized such that $\int P(\Delta U)d\Delta U=1$. The 
distributions decay exponentially $P(\Delta U) \sim \exp(-\kappa \Delta U)$ 
for all $R$ above and below the yielding transition. 
The dashed lines give least-square linear fits for each $R$. (c) 
The coefficient $\kappa$ of the exponential decay of $P(\Delta U)$ versus strain for the same 
cooling rates in (a) and (b). The inset shows the 
scaled coefficient $\widetilde \kappa$ of the exponential decay versus $\gamma$.}
\label{fig:dist}
\end{center}
\end{figure}

The above results for the rearrangement-induced energy drops were
obtained by ensemble averaging over many independent samples at each
strain and cooling rate. We will now consider the distribution of
energy drops as a function of strain and cooling rate. There have been
a number of prior studies of the distribution of rearrangements,
spanning length scales from avalanches in earthquakes and other
geophysical flows~\cite{bak1989earthquakes,kawamura2012statistical},
particle rearrangements in driven granular matter~\cite{denisov},
serrated flows in bulk metallic glasses~\cite{antonaglia2014bulk}, and
thermally activated particle rearrangements in amorphous
alloys~\cite{fan2015crossover}. The distribution of energy drops can
display power-law scaling or exponential decay depending on the
temperature and whether the driving is inertial or
overdamped~\cite{antonaglia2014tuned,fan2015crossover,sun2010plasticity,
  salerno2012avalanches}.  For amorphous systems with AQS driving, the form of the distribution of energy drops is
typically
exponential~\cite{maloney2006amorphous,lerner2009locality,maloney2004subextensive}.

In contrast to prior studies, we will characterize the form of the
probability distribution $P(\Delta U)$ of energy drops $\Delta U$ for each
rearrangement both before and after the yielding transition.  In
Fig.~\ref{fig:dist}, we show $P(\Delta U)$ from rearrangements before
yielding ($\gamma<0.055$ in panel (a)) and after yielding
($\gamma>0.055$ in panel (b)).  Both before and after yielding, the
probability distribution decays exponentially:
\begin{equation}
\label{eq:exp_dist}
P(\Delta U) = \frac{1}{\kappa} \exp(-\kappa \Delta U),
\end{equation}
where $\kappa$ is a function of both strain $\gamma$ and cooling rate $R$.
Before the yielding transition, $\kappa$ depends strongly on cooling rate,
{\it i.e.} $\kappa$ increases as the cooling rate decreases.  Slowly cooled
glasses have a relatively low probability for rearrangements with
large $\Delta U$ before yielding. After yielding, the distribution of
energy drops is only weakly dependent on cooling rate. In
Fig.~\ref{fig:dist} (c), we plot the coefficient $\kappa$ of the
exponential decay of the energy drop distribution $P(\Delta U)$ as a
function of strain $\gamma$ for several cooling rates $R$.  For more
slowly cooled glasses, there is a more rapid decrease in $\kappa$ before
yielding.  After yielding, $\kappa \simeq 100$ is independent of $\gamma$
and $R$ and similar to values found in related studies of
rearrangements in sheared binary Lennard-Jones 
glasses~\cite{lerner2009locality,maloney2006amorphous}.  The behavior
of the energy scale $1/\kappa$ mirrors the behavior of the average
potential energy $\langle U(\gamma) \rangle$
(Fig.~\ref{fig:energy} (b)). We find that $\widetilde \kappa = \kappa(a \langle
U(\gamma)\rangle + u) \sim 1$, where $a \approx 0.1$ is a constant,
and $u/\langle U \rangle \ll 1$ and does not depend on $\gamma$ or
$R$. In the inset to Fig.~\ref{fig:dist} (c), we show $\widetilde \kappa$
as a function of $\gamma$ and $R$.

\begin{figure}
\begin{center}
\includegraphics[width=0.9\columnwidth]{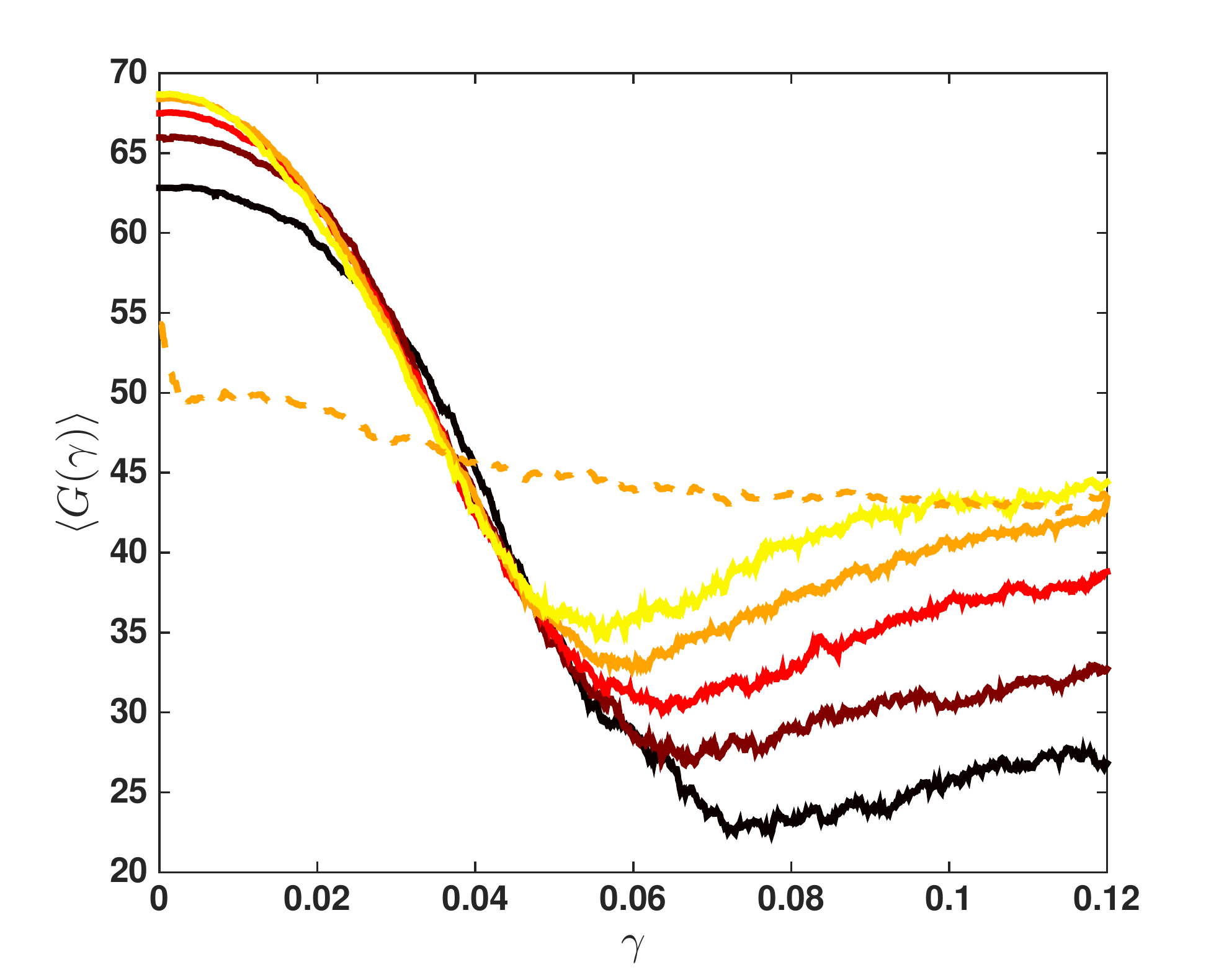}
\caption{The ensemble-averaged local slope
of the continuous stress versus strain segments $\langle G(\gamma) \rangle$ 
plotted versus strain $\gamma$ for glasses prepared with cooling rate 
$R=10^{-5}$ (solid curves)
and several system sizes: $N=250$ (black), $500$ (brown),
$1000$ (red), $2000$ (orange solid curve), and $4000$ (yellow). $\langle G(\gamma) \rangle$ for
rapidly cooled glasses with $R=10^{-1}$ and $N=2000$ is shown for 
comparison (dashed curve). All data have been averaged over at least 
$500$ samples.}
\label{fig:N_SS}
\end{center}
\end{figure}

We also studied the system-size dependence of the softening-induced
stress and energy losses.  The softening-induced stress loss per
strain $\langle d\sigma_{\rm loss}'(\gamma)/d\gamma\rangle$ is caused by decreases
in the local slopes of the continuous stress versus strain
segments. (See Fig.~\ref{fig:SS}.)  As the system size increases, the
frequency of rearrangements increases (as shown in
Fig.~\ref{fig:freqsize} (a)) and the lengths of the continuous
stress versus strain segments shorten. Here, we investigate whether the 
local slopes of the continuous stress versus strain segments change 
significantly with system size. 

In Fig.~\ref{fig:N_SS}, we show $\langle G(\gamma) \rangle$ for a slowly 
cooled glass ($R=10^{-5}$) as a
function of system size from $N=250$ to $4000$. $\langle G(\gamma)
\rangle$ is nearly independent of system size at strains prior to yielding
$\gamma \lesssim 0.055$.  In contrast, at large strains above
yielding, $\langle G(\gamma) \rangle$ grows (and $\langle d\sigma_{\rm
  loss}'/d\gamma\rangle$ decreases) with $N$. We see that for the larger
system sizes ($N > 1000$) $\langle G(\gamma)\rangle$ begins to saturate. For
comparison, we show $\langle G(\gamma) \rangle$ for a glass prepared
at the highest cooling rate studied, $R=10^{-1}$. At these cooling 
rates, $\langle \sigma(\gamma) \rangle$ reaches a large-strain plateau value 
that is only weakly system-size dependent. In Appendix~\ref{app:N}, we show
the system-size dependence of the softening-induced stress $\langle d\sigma_{\rm
  loss}'(\gamma)/d\gamma\rangle$ and energy $\langle dU_{\rm loss}'(\gamma)/d\gamma \rangle$
loss per strain.  Both quantities saturate in the large-system limit, with 
forms that are qualitatively the same as those for smaller system 
sizes.  Thus, softening-induced losses persist in the large-system 
limit. 

In this section, we presented results for the system-size dependence of the
rearrangement- and softening-induced stress and energy losses from
AQS pure shear as a function of strain and cooling rate.
Several quantities (both rearrangement- and softening-induced 
losses) show strong system-size dependence near yielding,
which serves to identify the onset of the transition from a solid-like
to a flowing state.  For example, the potential energy loss per strain
$\langle dU_{\rm loss}/d\gamma \rangle$ from rearrangements shows a sigmoidal form
that becomes increasingly sharp in the large-system limit and the
stress loss per strain $\langle d\sigma_{\rm loss}'/d\gamma \rangle$ from softening
shows significant system size dependence above yielding, but not
below.

\section{Conclusions and Future Directions} 
\label{sec:conclusions}

In this article, we characterized the nonlinear mechanical response of
binary Lennard-Jones glasses subjected to AQS pure
shear. We performed comprehensive numerical simulations as a function
of strain $\gamma$ above and below the yielding transition, cooling
rates $R$ used to prepare the zero-temperature glasses over five
orders of magnitude, and system sizes ranging from $N=250$ to $4000$.

To investigate the mechanical response, we focused on the von Mises
stress $\sigma$ and total potential energy per particle $U$. Though it
is hidden when taking an ensemble average, $\sigma(\gamma)$ and
$U(\gamma)$ for each single glass configuration are composed of
continuous segments in strain punctuated by rapid drops in either
stress or energy caused by particle rearrangements.  Thus, deviations
(typically losses) in the stress or potential energy from elastic
behavior originate from two sources: 1) softening-induced losses from
changes in the form of the continuous segments in strain and 2)
rearrangement-induced losses that depend on the frequency and size of
the energy or stress drops.  A key feature of this study is that we
decomposed the total stress and energy losses into contributions from
both sources.

In general, both softening- and rearrangement-induced losses are small
well below the yield strain, and then they begin to increase rapidly
near yielding.  Near and above yielding, both types of losses
contribute to the nonlinear mechanical response and remain finite in
the large-system limit. In the range of cooling rates studied here,
rearrangement-induced stress losses are larger than softening-induced
stress losses.  However, the softening-induced stress losses increase with
decreasing cooling rate (Fig.~\ref{fig:stress} (a)), and thus
softening-induced stress losses can dominate the nonlinear mechanical
response at sufficiently small cooling rates.

In many cases, the yield strain, where sheared glasses transition from
a disordered solid into a flowing state, is difficult to pinpoint
because many physical quantities, such as the shear stress and potential
energy, vary smoothly with strain~\cite{regev2013onset}. Here, we identified several
quantities that show significant changes as the strain is increased
above yielding. First, geometric features ({\it i.e.} the half-width
$W$ and depth $D$) of basins in the PEL along
the strain direction develop peaks near the yield strain for slowly
cooled glasses. In addition, the scaling relation between the
half-width and depth $D\sim W^{\lambda}$ changes from a scaling
exponent of $\lambda = 2$ below yielding to $1.5$ above yielding for
all cooling rates studied.  Second, the rearrangement-induced energy
loss per strain $dU_{\rm loss}/d\gamma$ possesses a sigmoidal form
with a midpoint near the yield strain that becomes sharper as the cooling rate
decreases and system size increases. Further, we decomposed the
rearrangement-induced energy loss per strain $\langle dU_{\rm loss}/d\gamma \rangle$
into two terms that determine the size and frequency of
rearrangements, and showed that the system-size scaling of these two
terms changes near the yielding transition~\cite{hentschel2015stochastic}.  Third, as found
previously, the distribution of energy drops decays exponentially for
AQS sheared glasses over the full range of
strain~\cite{maloney2006amorphous,lerner2009locality,maloney2004subextensive}.
However, the energy scale of the exponential decay depends
strongly on the cooling rate below yielding, while it is cooling-rate
independent above yielding.
 
In future studies, we will investigate several key open questions.
First, the current computational studies were performed using AQS pure
shear~\cite{maloney2006amorphous}.  How will the results we presented
change when we consider glasses sheared at finite shear rate $\dot
\gamma$ and temperature $T$?  Suppose the timescale for structural
relaxation from thermal fluctuations is given by $\tau$. In the case
$\dot \gamma \tau \ll 1$, we expect similar results to those presented
here.  As the temperature increases, the system will sample higher
regions of the PEL than sampled at zero temperature.  The frequency of
particle rearrangements will increase for $T>0$ as rearrangements
become thermally activated instead of strain-induced mechanical
instabilities~\cite{lemaitre2009rate,karmakar2010statistical2}.  In
future studies, we will analyze the rearrangement- and
softening-induced losses at finite temperate and strain rate to
determine their effects on the stress versus strain
curve~\cite{dubey2016elasticity,rottler2003shear}, yield
strain~\cite{johnson2005universal}, and
ductility~\cite{yu2012tensile}.

Second, the computational studies presented here were performed using
strain control, and thus at each strain, the system was mechanically
stable with a non-zero shear modulus.  In contrast, when sheared at
fixed shear stress, the system will flow with zero shear modulus until
the system finds a glass configuration with a shear stress that
matches the applied shear stress~\cite{dailidonis2014mechanical,lin2015criticality}.
If the system cannot find a
configuration that can balance the applied shear stress, the system
will flow indefinitely with a well-defined average shear rate.  In
future studies, we will compare the rearrangement- and
softening-induced stress and energy losses in the fixed shear stress
and strain ensembles.

In previous computational studies, we showed that sheared frictionless
granular materials, which interact via purely repulsive interactions,
possess monotonic stress versus strain curves even for ``slowly cooled''
granular samples~\cite{xu2006measurements}. Based on our current
results for rapidly cooled binary Lennard-Jones glasses, we expect that the stress and
potential energy losses in frictionless granular materials are
dominated by rearrangement-induced losses. In future studies, we will
determine the relative contributions of rearrangement- and
softening-induced stress and energy losses as a function of the
strength and range of the attractive interactions in the interatomic
potential.  In particular, recent studies~\cite{shi2014intrinsic,dauchot2011athermal}
have shown that the form of
the interaction potential can influence the ductility of amorphous
alloys, and thus we will investigate the relative contributions of
rearrangement- and softening-induced losses in ductile versus brittle
glasses.
 
\begin{acknowledgments}
The authors acknowledge primary financial support from NSF MRSEC
DMR-1119826 (K.Z.) and partial support from NSF Grant Nos.
CMMI-1462439 (C.O. and M.F.) and CMMI-1463455 (M.S.). 
This work was supported by the High Performance Computing facilities 
operated by, and the staff of, the Yale Center for Research Computing. 
\end{acknowledgments}

\appendix
\section{System-size scaling}
\label{app:N}

\begin{figure}
\begin{center}
\includegraphics[width=0.9\columnwidth]{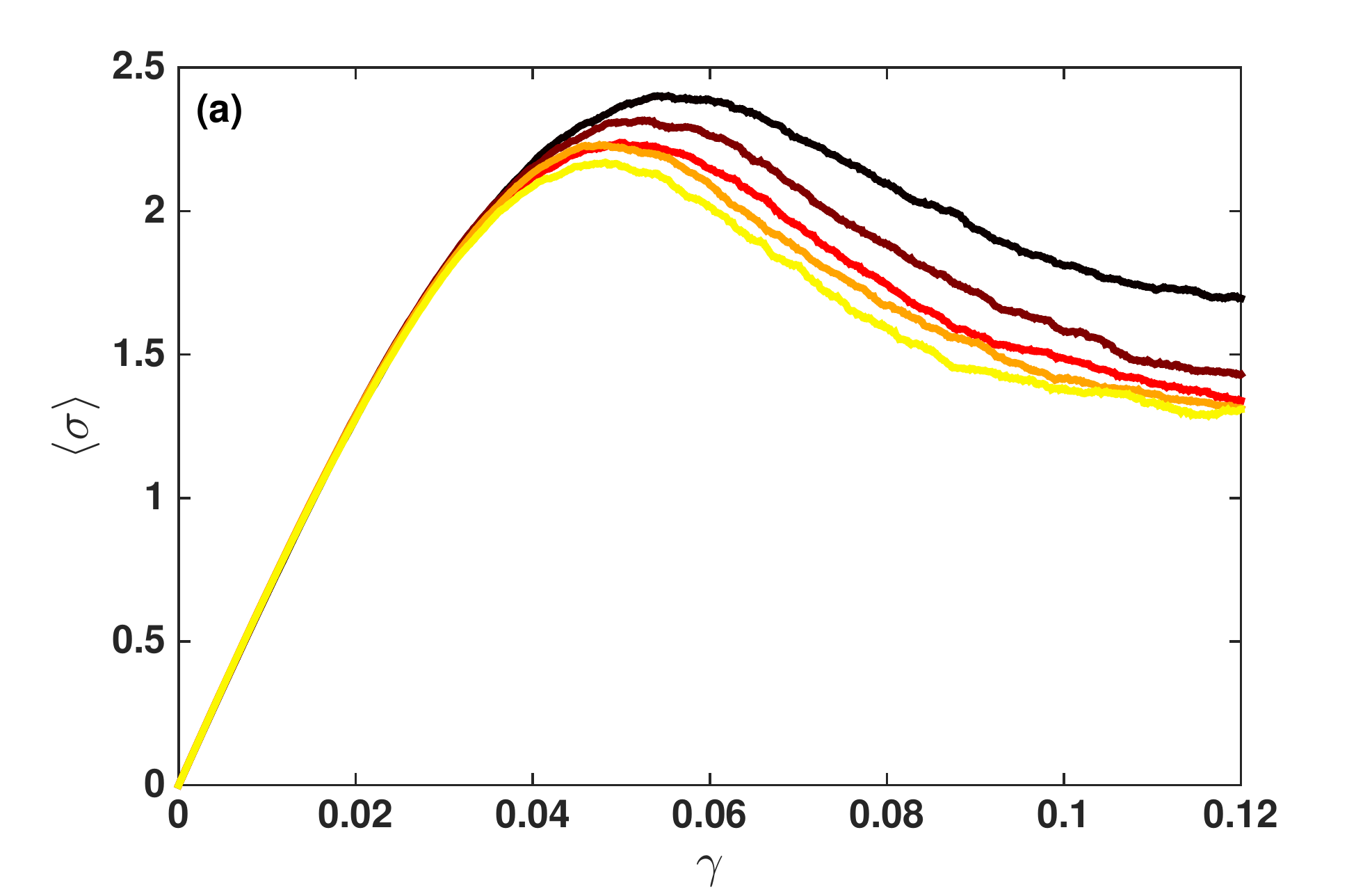}
\includegraphics[width=0.9\columnwidth]{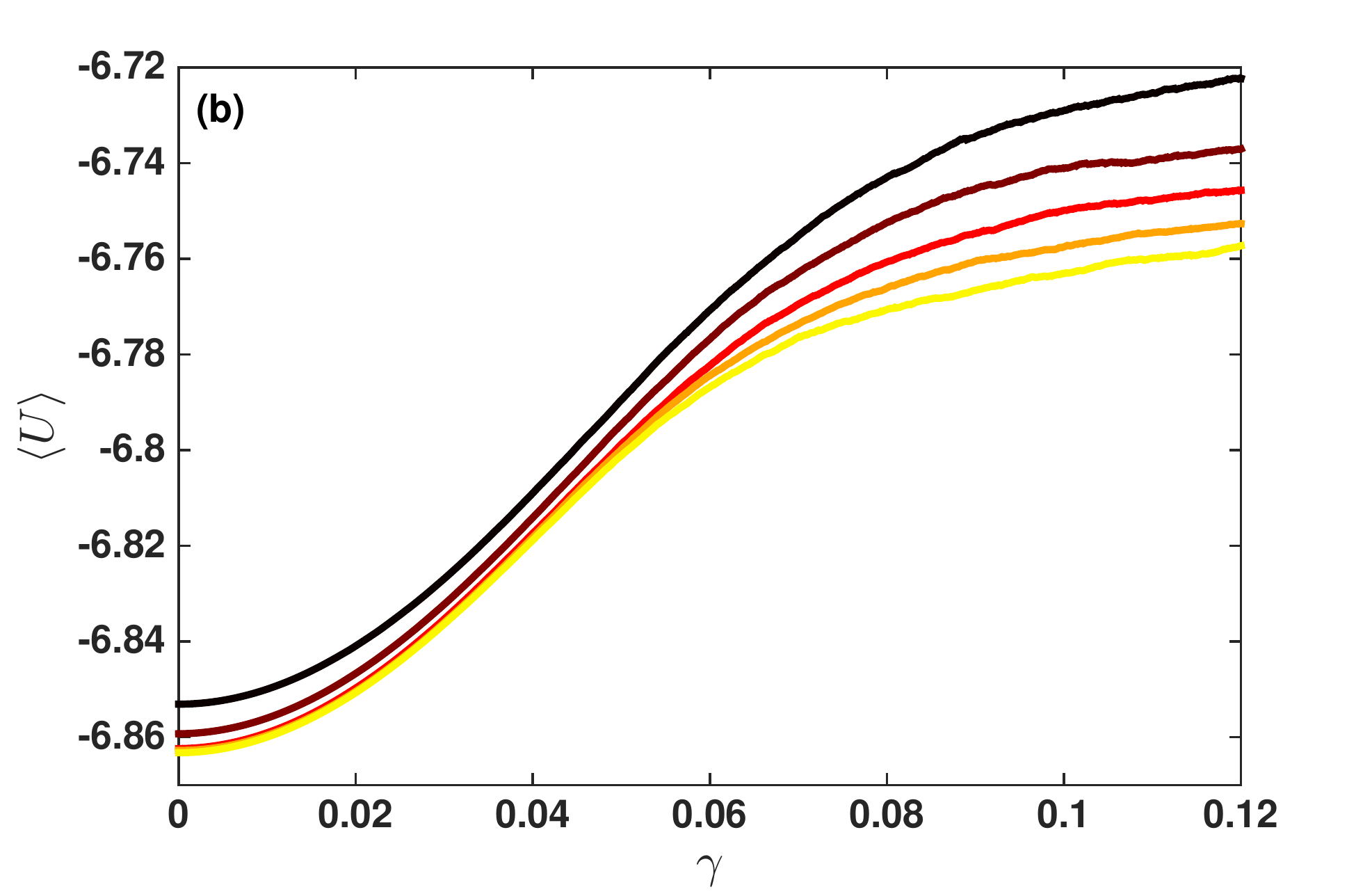}
\caption{The ensemble-averaged (a) von Mises stress $\langle \sigma\rangle$ and (b)
potential energy per particle $\langle U\rangle$ versus strain $\gamma$ for
systems prepared at cooling rate
$R=10^{-5}$ and several system sizes: $N=250$ (black), $500$
(brown), $1000$ (red), $2000$ (orange), and $4000$ (yellow). $\langle \sigma \rangle$ 
and $\langle U\rangle$ were averaged over at least $500$ independent samples.}
\label{fig:app_N1}
\end{center}
\end{figure}

In Figs.~\ref{fig:stress} and~\ref{fig:energy} in the main text, we
showed the ensemble-averaged von Mises stress $\langle \sigma\rangle$ and potential
energy per particle $\langle U\rangle$ versus strain for a single system size,
$N=2000$.  In Fig.~\ref{fig:app_N1}, we show $\langle \sigma(\gamma)\rangle$ and
$\langle U(\gamma)\rangle$ for system sizes ranging from $N=250$ to $4000$.  For
large $N > 1000$, $\langle \sigma(\gamma)\rangle$ and $\langle U(\gamma)\rangle$ appear to be approaching 
their large-system limits, although the system-size dependence at large 
strains is stronger than that at small strains.  

\begin{figure}
\begin{center}
\includegraphics[width=0.9\columnwidth]{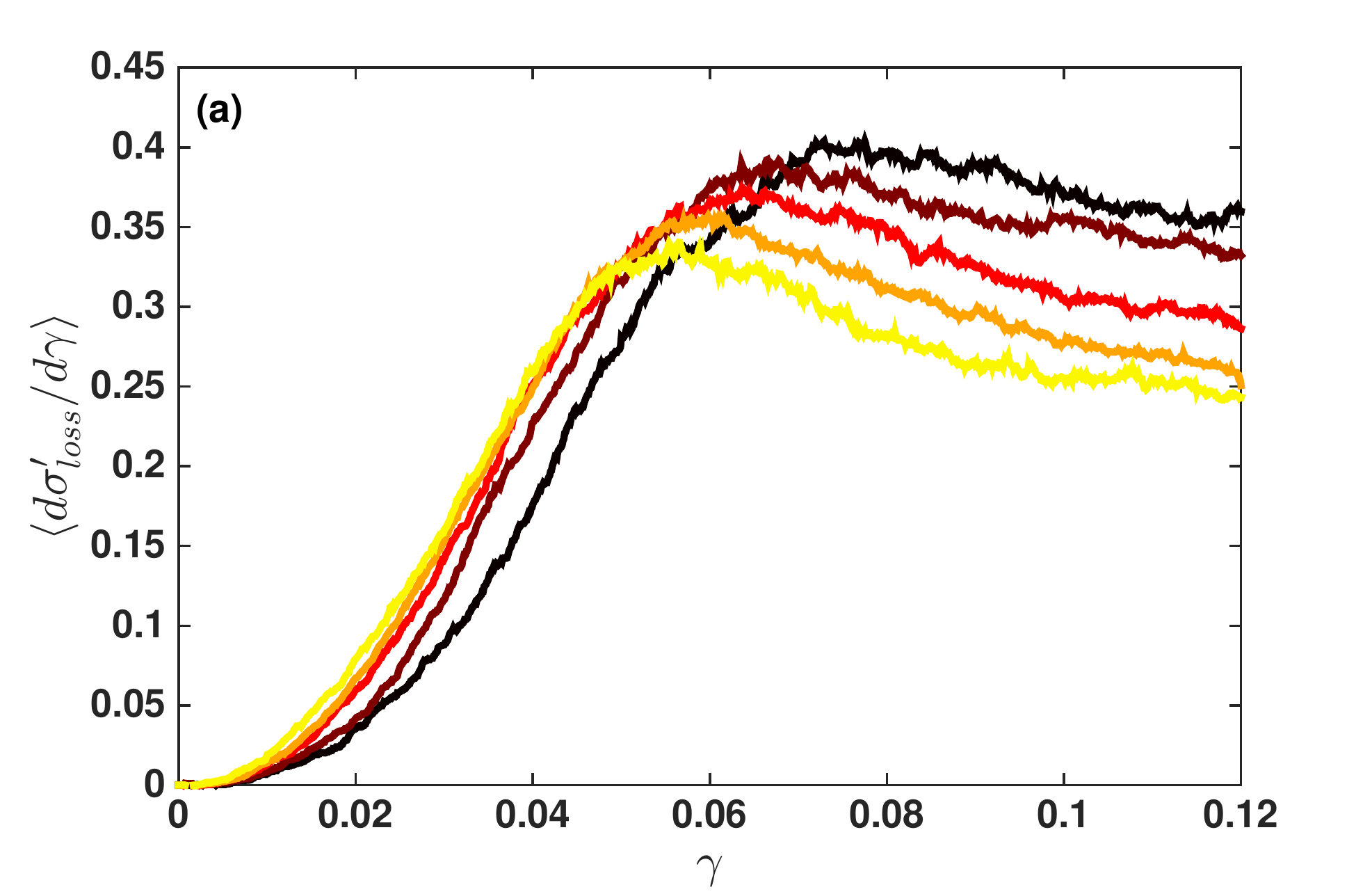}
\includegraphics[width=0.9\columnwidth]{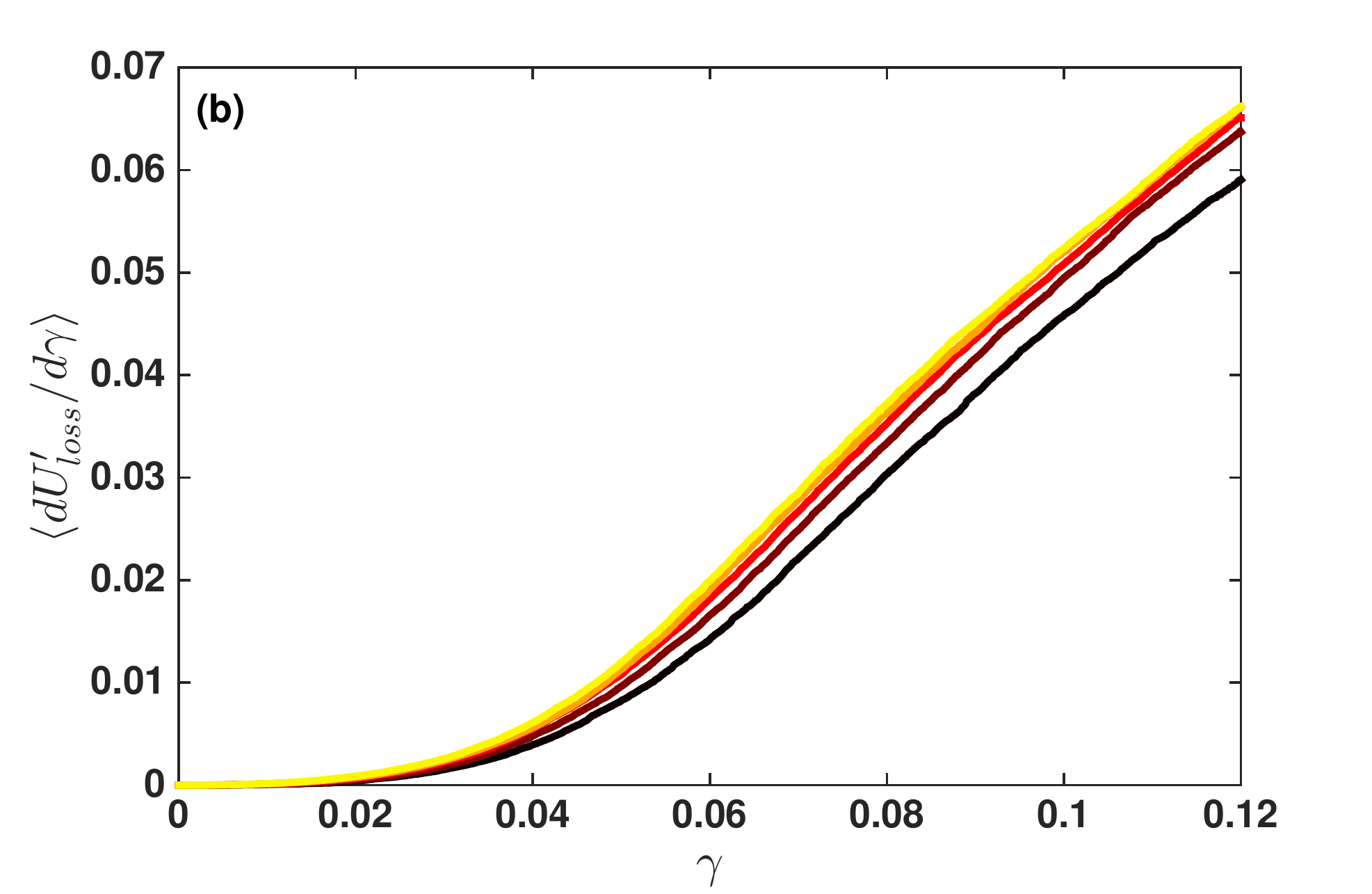}
\caption{The ensemble-averaged softening-induced (a)
stress loss per (1\%) strain $\langle d\sigma_{\rm
 loss}'/d\gamma\rangle$ and (b) energy loss per (1\%)
strain $\langle dU_{\rm loss}'/d\gamma\rangle$ for a slow cooling rate $R=10^{-5}$
and several system sizes: $N=250$ (black), $500$ (brown), $1000$
(red), $2000$ (orange), and $4000$ (yellow). $\langle d\sigma_{\rm loss}'/d\gamma\rangle$ 
and $\langle dU_{\rm loss}'/d\gamma\rangle$ were
averaged over at least $500$ independent samples.}
\label{fig:app_N2}
\end{center}
\end{figure}

In Fig.~\ref{fig:app_N2}, we show the ensemble-averaged
softening-induced stress loss per (1\%) strain $\langle d\sigma_{\rm
  loss}'/d\gamma\rangle$ and energy loss per (1\%) strain $\langle dU_{\rm
  loss}'/d\gamma\rangle$. For $N \gtrsim 1000$, $\langle dU_{\rm loss}'/d\gamma\rangle$ is nearly
independent of system size. Below yielding ($\gamma \lesssim 0.055$),
$\langle d\sigma_{\rm loss}'/d\gamma\rangle$ reaches its large-system limiting form
for $N \gtrsim 1000$.  In contrast, above yielding, $\langle d\sigma_{\rm
  loss}'/d\gamma\rangle$ has stronger system-size dependence, although it
appears that $\langle d\sigma_{\rm loss}'/d\gamma\rangle$ will saturate in the
large-system limit.  This behavior is similar to that found for the
local shear modulus $\langle G(\gamma) \rangle$ in
Fig.~\ref{fig:N_SS}.

\bibliography{nonlinearity}

\end{document}